\documentclass[pdflatex,sn-standardnature]{revtex4-1}
\usepackage{pdfpages}
\usepackage{etoolbox} 
\usepackage{float}
\usepackage{graphicx}
\usepackage{dcolumn}

\usepackage{bm}
\usepackage{adjustbox}
\usepackage[utf8]{inputenc}
\usepackage[T1]{fontenc}
\usepackage{mathptmx}
\usepackage{xcolor}
\usepackage{chemformula}
\usepackage{orcidlink}%
\usepackage{hyperref}
\hypersetup{
	colorlinks   = true,
	citecolor    = blue,
	linkcolor = blue,
    urlcolor = blue
}
\makeatletter
\patchcmd{\@outputpage@head}{\@ifx{\LS@rot\@undefined}{}{\LS@rot}}{}{}{}
\makeatother

\def\checkmark{\tikz\fill[scale=0.4](0,.35) -- (.25,0) -- (1,.7) -- (.25,.15) -- cycle;}

\begin{document}



\title{JARVIS-Leaderboard: A Large Scale Benchmark of Materials Design Methods}
\author{Kamal Choudhary\orcidlink{0000-0001-9737-8074}}
 \email{kamal.choudhary@nist.gov}

 \affiliation{ 
Material Measurement Laboratory, National Institute of Standards and Technology, Maryland, 20899, USA.
}%

\author{Daniel Wines\orcidlink{0000-0003-3855-3754}}%

\affiliation{ 
Material Measurement Laboratory, National Institute of Standards and Technology, Maryland, 20899, USA.
}%

\author{Kangming Li\orcidlink{0000-0003-4471-8527}
}
\affiliation{Department of Materials Science and Engineering, University of Toronto, 27 King's College Cir, Toronto, ON, Canada.}

\author{Kevin F. Garrity\orcidlink{0000-0003-0263-4157}}

\affiliation{ 
Material Measurement Laboratory, National Institute of Standards and Technology, Maryland, 20899, USA.
}%

\author{Vishu Gupta\orcidlink{0000-0002-4931-7194}}

\affiliation{ 
Department of Electrical and Computer Engineering, Northwestern University, Evanston, Illinois, 60208, USA
}%

\affiliation{ 
Lewis-Sigler Institute for Integrative Genomics, Princeton University, Princeton, New Jersey, 08544, USA 
}%

\affiliation{ 
Ludwig Institute for Cancer Research, Princeton University, Princeton, New Jersey, 08544, USA 
}%

\author{Aldo H. Romero\orcidlink{0000-0001-5968-0571}}

\affiliation{ 
Department of Physics and Astronomy, West Virginia University, Morgantown, WV 26506, USA
}%

\author{Jaron T. Krogel\orcidlink{0000-0002-1859-181X}}

\affiliation{ 
Materials Science and Technology Division, Oak Ridge National Laboratory, Oak Ridge,Tennessee 37831,USA
}%

\author{Kayahan Saritas\orcidlink{0000-0002-2240-8520}}

\affiliation{ 
Materials Science and Technology Division, Oak Ridge National Laboratory, Oak Ridge,Tennessee 37831,USA
}%

\author{Addis Fuhr\orcidlink{0000-0002-8819-8255}}

\affiliation{ 
Center for Nanophase Materials Science, Oak Ridge National Laboratory, Oak Ridge, Tennessee 37831, United States
}%

\author{Panchapakesan Ganesh\orcidlink{0000-0002-7170-2902}}

\affiliation{ 
Center for Nanophase Materials Science, Oak Ridge National Laboratory, Oak Ridge, Tennessee 37831, United States
}%

\author{Paul R. C. Kent\orcidlink{0000-0001-5539-4017}}

\affiliation{ 
Computational Sciences and Engineering Division, Oak Ridge National Laboratory, Oak Ridge, TN, 37831, USA
}%





\author{ Keqiang Yan\orcidlink{0009-0009-9286-9259}}

\affiliation{ 
Department of Computer Science and Engineering, Texas A\&M University, College Station, TX 77843, USA
}%

\author{Yuchao Lin\orcidlink{0009-0003-6271-6397}}

\affiliation{ 
Department of Computer Science and Engineering, Texas A\&M University, College Station, TX 77843, USA
}%

\author{Shuiwang Ji\orcidlink{0000-0002-4205-4563}}

\affiliation{ 
Department of Computer Science and Engineering, Texas A\&M University, College Station, TX 77843, USA
}%

\author{Ben Blaiszik\orcidlink{0000-0002-5326-4902}}

\affiliation{ 
Globus, University of Chicago, Illinois, 60637, USA.
Data Science and Learning Division, Argonne National Lab, Illinois, 60439, USA.
}%

\author{Patrick Reiser\orcidlink{0000-0002-7052-696X}}
\affiliation{ 
Institute of Nanotechnology, Karlsruhe Institute of Technology, Kaiserstraße 12, 76131 Karlsruhe, Germany
}%

\author{Pascal Friederich\orcidlink{0000-0003-4465-1465}}

\affiliation{ 
Institute of Theoretical Informatics, Karlsruhe Institute of Technology, Kaiserstraße 12, 76131 Karlsruhe, Germany
}%
\affiliation{ 
Institute of Nanotechnology, Karlsruhe Institute of Technology, Kaiserstraße 12, 76131 Karlsruhe, Germany
}%

\author{Ankit Agrawal\orcidlink{0000-0002-5519-0302}}

\affiliation{ 
Department of Electrical and Computer Engineering, Northwestern University, Evanston, Illinois, 60208, USA
}%
\author{Pratyush Tiwary\orcidlink{0000-0002-2412-6922}}

\affiliation{ 
Department of Chemistry and Biochemistry and Institute for Physical Science and Technology, University of Maryland, College Park, MD 20742, United States
}%

\author{Eric Beyerle\orcidlink{}}

\affiliation{ 
Department of Chemistry and Biochemistry and Institute for Physical Science and Technology, University of Maryland, College Park, MD 20742, United States
}%

\author{Peter Minch\orcidlink{}}

\affiliation{ 
Department of Physics, Applied Physics and Astronomy, Rensselaer Polytechnic Institute, Troy, NY 12180, USA
}%

\author{Trevor David Rhone\orcidlink{0000-0002-0198-9952}}

\affiliation{ 
Department of Physics, Applied Physics and Astronomy, Rensselaer Polytechnic Institute, Troy, NY 12180, USA
}%

\author{Ichiro Takeuchi\orcidlink{0000-0002-1366-1946}}

\affiliation{ 
Department of Materials Science and Engineering, University of Maryland, College Park, MD 20742, USA
}%
\author{Robert B. Wexler\orcidlink{0000-0002-6861-6421}
}
\affiliation{Department of Chemistry and Institute of Materials Science and Engineering, Washington University in St. Louis, St. Louis, MO 63130, USA}

\author{Arun Mannodi-Kanakkithodi\orcidlink{0000-0003-0780-1583}}

\affiliation{ 
School of Materials Engineering, Purdue University, West Lafayette, IN, 47907, USA
}%

\author{Elif Ertekin\orcidlink{0000-0002-7816-1803}}

\affiliation{ 
Department of Mechanical Science and Engineering, University of Illinois Urbana-Champaign, Urbana, Illinois 61801, USA
}%
\affiliation{Materials Research Laboratory, University of Illinois Urbana-Champaign, Urbana, Illinois 61801, USA}

\author{Avanish Mishra\orcidlink{0000-0003-3997-0445}}
\author{Nithin Mathew\orcidlink{0000-0002-2316-3190}}

\affiliation{ 
Theoretical Division (T-1), Los Alamos National Laboratory, Los Alamos, NM, 87545, USA
}%



\author{Mitchell Wood\orcidlink{0000-0001-5878-4096}}

\affiliation{ 
Center for Computing Research, Sandia National Laboratories, Albuquerque, New Mexico 87185, USA
}%

\author{Andrew Dale Rohskopf\orcidlink{0000-0002-2712-8296}}

\affiliation{ 
Center for Computing Research, Sandia National Laboratories, Albuquerque, New Mexico 87185, USA
}

\author{Jason Hattrick-Simpers\orcidlink{0000-0003-2937-3188}
}
\affiliation{Department of Materials Science and Engineering, University of Toronto, 27 King's College Cir, Toronto, ON, Canada.}



\author{Shih-Han Wang \orcidlink{0000-0003-4418-2080}}
\affiliation{Department of Chemical Engineering, Virginia Polytechnic Institute and State University, Blacksburg, VA 24061, USA}

\author{Luke E.~K.~Achenie \orcidlink{0000-0001-9850-5346}}
\affiliation{Department of Chemical Engineering, Virginia Polytechnic Institute and State University, Blacksburg, VA 24061, USA}

\author{Hongliang Xin \orcidlink{0000-0001-9344-1697}}
\affiliation{Department of Chemical Engineering, Virginia Polytechnic Institute and State University, Blacksburg, VA 24061, USA}





\author{Maureen Williams\orcidlink{0000-0001-9144-0551}}%

\affiliation{ 
Material Measurement Laboratory, National Institute of Standards and Technology, Maryland, 20899, USA.
}%

\author{Adam J. Biacchi\orcidlink{0000-0001-5663-2048}}

\affiliation{ 
Physical Measurement Laboratory, National Institute of Standards and Technology, Maryland, 20899, USA.
}%

\author{Francesca Tavazza\orcidlink{0000-0002-5602-180X}}

\affiliation{ 
Material Measurement Laboratory, National Institute of Standards and Technology, Maryland, 20899, USA.
}%



\begin{abstract}

Lack of rigorous reproducibility and validation are significant hurdles for scientific development across many fields. Materials science, in particular, encompasses a variety of experimental and theoretical approaches that require careful benchmarking. Leaderboard efforts have been developed previously to mitigate these issues. However, a comprehensive comparison and benchmarking on an integrated platform with multiple data modalities with perfect and defect materials data is still lacking. This work introduces JARVIS-Leaderboard, an
open-source and community-driven platform that facilitates benchmarking and enhances reproducibility. The platform allows
users to set up benchmarks with custom tasks and enables contributions in the form of dataset, code, and meta-data submissions. We
cover the following materials design categories: Artificial Intelligence (AI), Electronic Structure (ES), Force-fields (FF),
Quantum Computation (QC) and Experiments (EXP). For AI, we cover several types of input data, including atomic structures,
atomistic images, spectra, and text. For ES, we consider multiple ES approaches, software
packages, pseudopotentials, materials, and properties, comparing results to experiment. For FF, we compare multiple approaches for material property predictions. For QC, we benchmark Hamiltonian simulations using various quantum algorithms and circuits. Finally, for experiments, we use the inter-laboratory approach to establish benchmarks.
There are 1281 contributions to 274 benchmarks using 152 methods with more than 8 million data-points, and the leaderboard is continuously
expanding. The JARVIS-Leaderboard is available at the website: \url{https://pages.nist.gov/jarvis_leaderboard/}




\end{abstract}

\maketitle


\section{Introduction}

The accelerated design and characterization of materials of technological interest has been a rapidly evolving area of research in the last few decades
\cite{ward2015materials}. Materials design requires approaches spanning a variety of length and time
scales\cite{callister2000fundamentals}. For atomistic design, the methods employed may include computational approaches such as density functional theory,
tight-binding, force-fields, and highly accurate approaches such as quantum Monte Carlo or quantum computations. A wide range of approaches are employed above the purely atomistic level, such as mesoscale and finite-element methods. Similarly, experimental characterization approaches include X-ray diffraction, vibroscopy, manometry, scanning electron microscopy, and magnetic susceptibility measurements. 



Moreover, data produced from
these techniques can be of various types: chemical formulae, atomic/micro-structures, images, spectra, and
text-documents\cite{agrawal2020materials,choudhary2022recent,doi:10.1142/9789811265679_0023}. The data analysis and curation methods add further
complexity to benchmarking efforts, which are extremely important \cite{camerer2018evaluating,fanelli2018reproducibility,sun2020evaluating,amrhein2017earth,grimes2018modelling,allen2023interpretable,prager2019improving,papadiamantis2020metadata,zhu2023reproducibility,lehtola2023reproducibility,sayre2018reproducibility,papadiamantis2024metadata}. For example, more than 70 \% of research works were shown to be non-reproducible \cite{park2017reproducible,baker2016reproducibility,hutson2018ai}, and this number could be much higher depending upon the field of investigation. Although there have been
significant advances in individual fields, there is an urgent need to establish a large-scale benchmark for systematic,
reproducible, transparent, and unbiased scientific development . 

Developing such metrology is a highly challenging task, even for
one of these methods, let alone the entire galaxy of available methods. Projects and approaches such as the materials genome and FAIR initiatives
\cite{ward2015materials,wilkinson2016fair}, have resulted in several well-curated
datasets and benchmarks. These, in turn, have led to several materials informatics applications\cite{agrawal2016perspective,RICKMAN2019473, agrawal2019deep, gupta2023evolution}. Although electronic structure approaches such as density functional theory (DFT) tend to be more reproducible than other categories \cite{lejaeghere2016reproducibility,lehtola2023reproducibility}, a systematic effort must be made to validate these methods and estimate the error in predictions. Hence, it is highly desirable to have a large-scale benchmarking platform in the materials science field for reproducibility and method validation. 

Massive progress in fields such as image
recognition/image classification (ImageNet \cite{russakovsky2015imagenet}), protein structure prediction (AlphaFold \cite{jumper2021highly}),  large language modeling (Generative pretrained transformers (GPT)) \cite{brown2020language}) has been possible primarily because of well-defined benchmarks in respective fields. With regards to AI methods for structure-to-property predictions \cite{zhang2023artificial}, benchmarking efforts have enabled drastic improvements in the accuracy of predicted properties (i.e., moving away from descriptor-based predictions and including graph neural networks in the model architectures to improve accuracy). 

For deterministic electronic structure methods such as DFT, extensive benchmarking of software and different DFT approximations (functionals, pseudopotentials, etc.) has led to increased reproducibility and precision in individual results and workflows \cite{dft-reprod,lejaeghere2016reproducibility}. Such benchmarks allow a wide community to solve problems collectively and systematically. In addition, since there already exists highly accurate models for specific tasks (i.e., energy prediction), more comprehensive evaluations of the models are required so that the performance ranking is not overfitted to one biased data source. We
believe that such a universal and large-scale set of benchmarks for materials science will significantly benefit the scientific
community.

To this date, several benchmarks of individual methods have already been developed. For artificial intelligence (AI) methods, there have been several
benchmarks and leaderboards such as MatBench \cite{dunn2020benchmarking}. \textcolor{black}{MatBench provides a leaderboard for machine learned structure-based property predictions of inorganic materials using 13 supervised machine learning tasks (thermodynamic, tensile, optical, thermal, elastic, and electronic properties) from 10 datasets (including DFT and experiment) \cite{dunn2020benchmarking}.} Similar AI benchmarking and leaderboard platforms include MoleculeNet \cite{wu2018moleculenet}, OpenCatalystProject\cite{chanussot2021open}, sGDML
\cite{chmiela2017machine,chmiela2019sgdml}, mLEARN \cite{zuo2020performance}, MatScholar \cite{weston2019named}, and AtomAI
\cite{ziatdinov2021atomai}. For electronic structure methods, some of the notable benchmarks include the work by Lejaeghere et
al.\cite{lejaeghere2016reproducibility}, Borlido et al.\cite{borlido2019large}, Huber et al. \cite{huber2021common}, Zhang et
al.\cite{zhang2018performance}, Tran et al.\cite{tran2022open} and several other projects
\cite{jurevcka2006benchmark,brauer2016s66x8,mata2017benchmarking,taylor2016blind}. Other method benchmarks include phase-field
benchmarks by Wheeler et al.\cite{wheeler2019pfhub}, Lindsay et al. \cite{lindsay2022moose}, and microscopy benchmarks such as by
Wei et al.\cite{wei2022benchmark}. \textcolor{black}{A few additional benchmarking studies in materials science include Refs. \onlinecite{Ren_2022_CVPR,8972912,HENDERSON2021107262,fung-benchmark,CECEN201876,BAIRD2023109487,CHEN2019109155,quartet,fu2022materials,C8ME00012C,LEJEUNE2020100659,aflow-benchmark,10.1063/5.0133528}. More details on some of these benchmarking efforts are provided in later sections.}

\textcolor{black}{The goal of this project is to provide a more comprehensive framework for materials benchmarking than previous works. In particular, most existing efforts: 1) lack the flexibility to readily incorporate new tasks or benchmarks, which is a limitation given the continuous discovery of new materials and quantities in science, 2) are specialized towards a single modality, such as electronic structure, rather than providing a comprehensive framework that can accommodate multiple modalities, 3) offer only a limited set of tasks or properties, 4) are primarily focused on computational methods, overlooking the importance of experimental benchmarking, and 5) make adding contributions to existing platforms rather complex, creating a barrier to entry. In general, there is a need to simplify the process of user contributions to leaderboards to foster broader community engagement.}


In this work, we present a user-friendly, comprehensive approach to integrate the benchmarking of both computational, experimental and
data-analytics methods. The JARVIS-Leaderboard framework (\url{https://pages.nist.gov/jarvis_leaderboard/}) covers a variety of
categories:  Artificial Intelligence (AI), Electronic Structure (ES), Force-field (FF), Quantum Computation (QC), and Experiments
(EXP). It also covers various data types, including atomic structures, spectra, images, and text. This project can be used to:
(1) check the state-of-the-art methods in respective fields, (2) add a contribution model on an existing benchmark, (3) add a new
benchmark, (4) compare new ideas and approaches to well-known approaches. To enhance reproducibility, we encourage each contribution
to (1) be from peer-reviewed articles with an associated DOI for all contributions, models, and tools, (2) include a run script to
exactly reproduce the results (especially for computational tools), (3) include a metadata file with details such as team name,
contact information, computational timing and software (with software version)/hardware used in order to enhance transparency. 

It is important to note differences between a typical data-repository and a benchmarking platform. Some of the key distinguishing factors between a usual large data-repository (such as JARVIS-DFT) and the present leaderboard effort are: 1) the leaderboard contains well-characterized/well-known samples/tasks (i.e., with digital object identifier/peer-reviewed article links) with all the scripts/metadata easily available to reproduce the results rather than just being a look-up table to find data, 2) large data repositories usually contain more variation in materials chemistry/structure and less variation of methods while the leaderboard focuses on a larger number of method comparisons. 

For example, the JARVIS-DFT contains DFT data for more than 80,000 materials and millions of material properties with a few specific ES methods and hence there are only a few entries for, say, the electronic bandgap of Silicon from different methods, while the leaderboard contains electronic bandgaps for Silicon using more than 17 ES methods from various contributors. Similarly, JARVIS-ALIGNN project contains AI models for more than 80 properties/tasks of materials, i.e., just one model for a well-known property such as formation energy, while there are more than 12 methods for formation energy task in the leaderboard (as discussed later).

Furthermore, the JARVIS-leaderboard attempts to bridge together multiple categories of methods (AI, ES, FF, QC, EXP) and types of data (single properties, structure, spectra, text, etc.) with the goal of broadening benchmarking efforts across several fields of study. What differentiates the JARVIS-Leaderboard from platforms such as MatBench \cite{dunn2020benchmarking}, is that MatBench \cite{dunn2020benchmarking} provides a handful of tasks to evaluate ML methods on larger datasets (i.e. 10$^4$ entries, most of which are from the Materials Project \cite{doi:10.1063/1.4812323}). A potential drawback of this approach is that the resulting performance rankings could be biased towards the data distribution of a single source. In contrast, the JARVIS-Leaderboard covers a broader range of datasets and properties and provides a better overview of model performance.

 
Recently in the field of machine learning in materials science, there has been a fixation on performance metrics for newly developed models. This begs the question of whether or not benchmarking can be destructive to the development of new methods if these new methods cannot immediately outperform the previous state-of-the-art approaches. This also begs the question of whether or not benchmarking can lead to overfitting or poor generalization \cite{li2023critical,li2023redundancy}. 

Therefore, we outline how the leaderboard can also be used to identify and focus on some of the major challenges in different fields, such as: (1) how to evaluate extrapolation capability\cite{choudhary2023can}? (2) why is it difficult to
 develop a reasonably good AI model with similar accuracy to electronic structure methods?, (3) how can we reduce the computational cost of higher accuracy electronic structure
 predictions (such as bandgaps and bandoffsets)?, (4) how do we identify examples of materials that require high-fidelity methods (beyond DFT accuracy)?, (5) how can we identify material space where methodological improvements need to be targeted?, (6) how can we establish figures of merits for mesoscale models such as phase field?, (7) how can we make atomistic image analysis quantitative rather than qualitative?, (8) and how do we develop and benchmark multi-modal models (such as text, image, video, atomic structures etc.) \cite{NEURIPS2019_e4da3b7f}?
 
The JARVIS-Leaderboard is seamlessly integrated into the existing and well-established NIST-JARVIS infrastructure \cite{choudhary2020joint,10.1063/5.0159299}, which hosts several
 datasets, tools, applications, and tutorials for materials design, motivated by the materials genome initiative \cite{ward2015materials}. The framework is open access to the entire materials science community for
 progressing the field collectively and systematically. JARVIS (Joint Automated Repository for Various Integrated Simulations) \cite{choudhary2020joint,10.1063/5.0159299}
 is a repository designed to automate materials discovery and optimization using classical force-field, density functional theory,
 machine learning calculations, and experiments. Nevertheless, the leaderboard is not limited to NIST-JARVIS infrastructure and can be linked with other external projects as well. 
 
 Since its creation in 2017, JARVIS has had over 50,000 users worldwide, over 45 JARVIS-associated articles have been published, and over 80,000  materials currently reside in the database. As these numbers continue to multiply, significant effort on external outreach to the materials science community has been an additional goal of JARVIS, with several events (\url{https://jarvis.nist.gov/events/}) such as the Artificial Intelligence for Materials Science (AIMS) and Quantum Matters in Materials Science (QMMS) workshops and hands-on JARVIS-Schools, which have had hundreds of participants throughout the last few years. Based on the level of success and support from the community with regard to the existing JARVIS infrastructure, we believe that the integration of the JARVIS-Leaderboard will have a similar level of engagement and success, with a growing number of contributors from all over the world (in government, academia and industry) and in different sub-fields of materials science.


\section{Results and discussion}
 \subsection{Leaderboard overview}

 \begin{figure*}[hbt!]
    \centering
    \includegraphics[trim={0. 0cm 0 0cm},clip,width=0.98\textwidth]{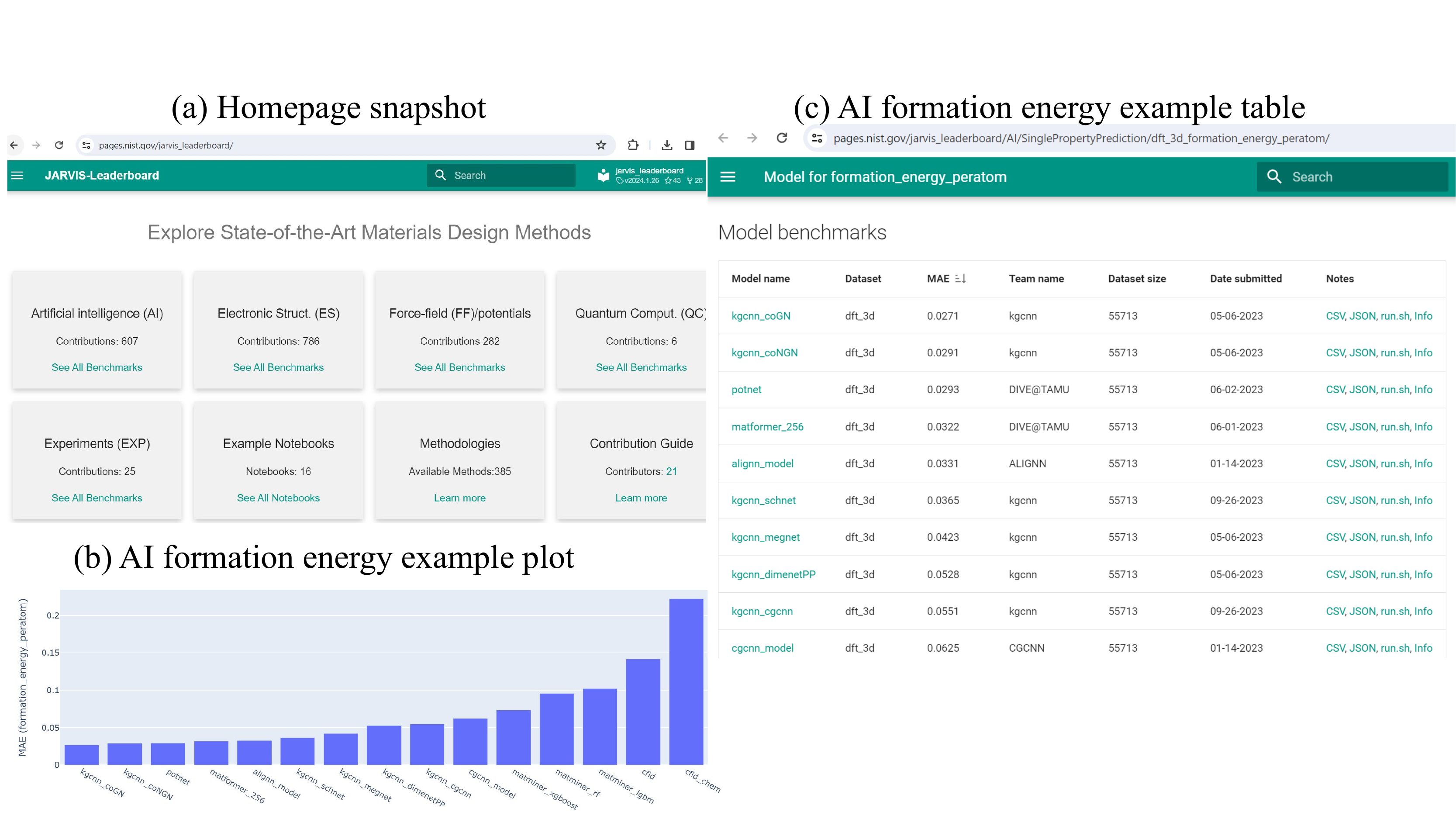}
    \caption{\label{fig:snapshot}{Leaderboard snapshot with an example output for AI based formation energy per atom model on the JARVIS-DFT (dft\_3d) dataset. a) homepage sanpshot showing list of categories and number of available contributions at the time of writing, b) an example AI regression model benchmark for formation energy with several contributions. The methods are sorted based on the mean absolute error (MAE) values. Lower MAE values indicate higher accuracy, c) explicit table for the plot in panel b.Links to 
 individual csv.zip (AI-SinglePropertyPrediction-formation\_energy\_peratom-dft\_3d-test-mae.csv.zip),  json.zip (dft\_3d\_formation\_energy\_peratom.json.zip), shell script (run.sh) and detailed info (metadata.json) files are provided to help enhance reproducibility. Such results plots and tables are available for each benchmark in the leaderboard.}}
\end{figure*}

 \begin{figure*}[hbt!]
    \centering
    \includegraphics[trim={0. 0cm 0 0cm},clip,width=0.98\textwidth]{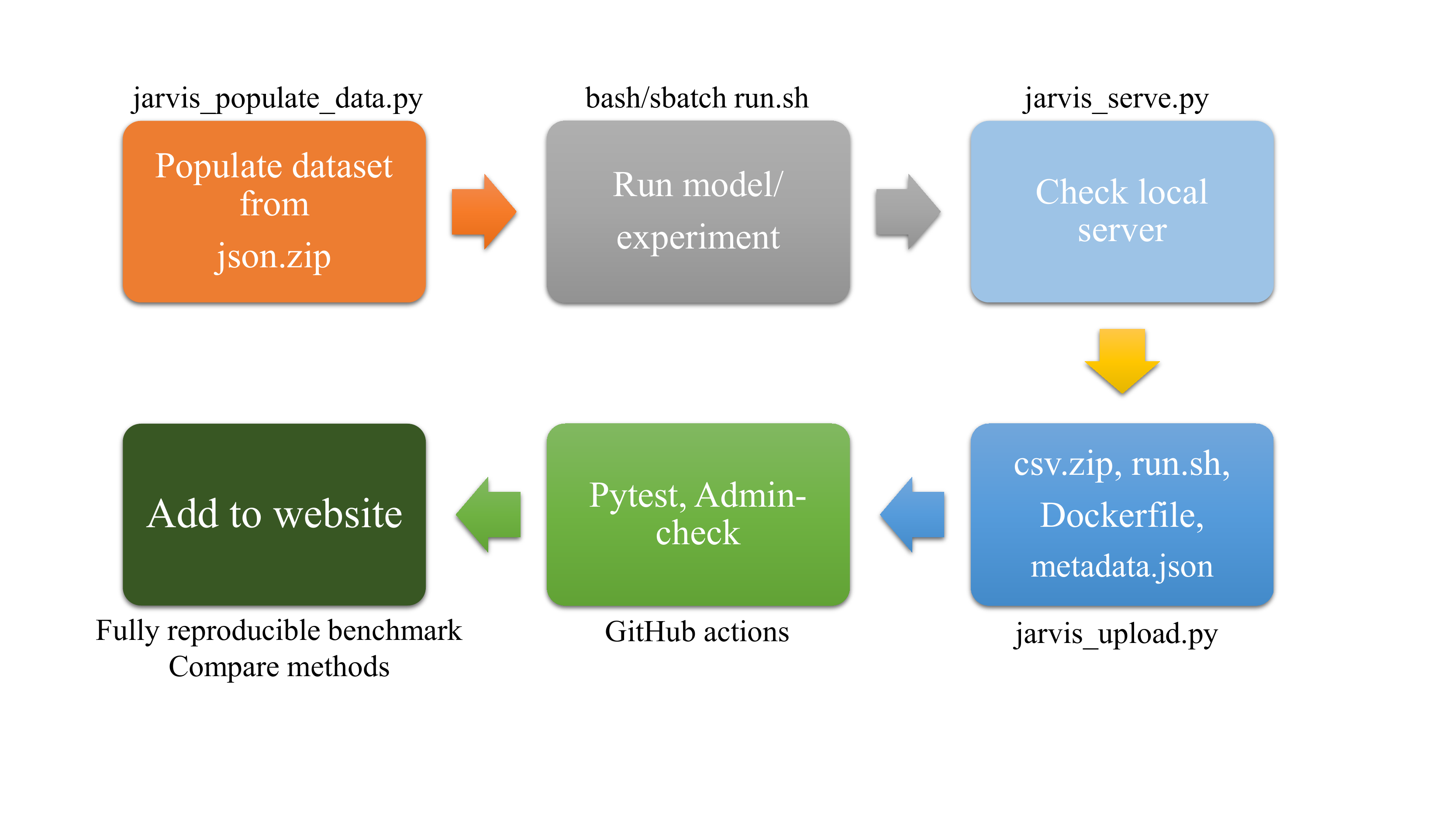}
    \caption{\label{fig:flowchart}{A flow-chart showing the processes involved in uploading a new contribution to the leaderboard. The jarvis\_populate\_data.py scripts generate a benchmark dataset. A user can apply their method, train
models, or run experiments on that dataset and prepare a csv.zip, a metadata.json file, and other files in a new folder in the contributions directory. The contributions can be locally checked by the user using jarvis\_server.py script. Then the folder can be uploaded to a user's GitHub account by the automated jarvis\_upload.py script involving several GitHub uploading steps. The administrators of the JARVIS-Leaderboard at NIST will verify the contributions and then finally, it will become part
of the leaderboard website.}}
\end{figure*}

\begin{figure*}[hbt!]
    \centering
    \includegraphics[trim={0. 0cm 0 0cm},clip,width=0.98\textwidth]{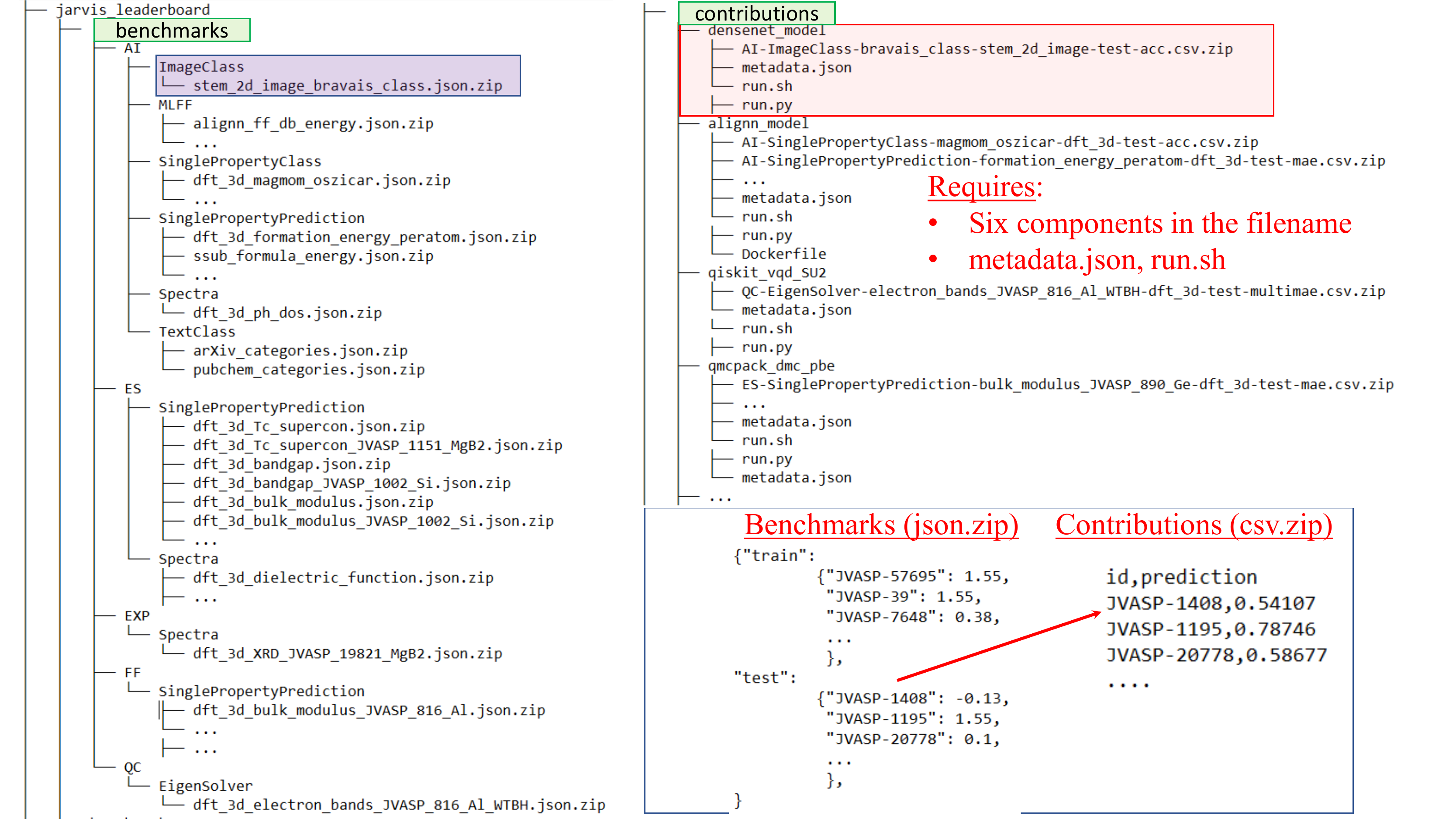}
    \caption{\label{fig:tree}{A tree diagram for directory and file-structure in the leaderboard. There are two main directories in the repo: (1) benchmarks (reference) and (2) leaderboard
contributions (for various leaderboard entries). In the ``benchmarks'' directory, there are folders for the AI, ES, QC, FF, and EXP categories. Within them, there are sub-folders for
specific sub-categories. In the ``contributions'' directory there is a collection of folders that consists of .csv.zip, metadata.json files, and
optionally a Dockerfile and run.sh file for available contributions from each method. The csv.zip file contains entries of identifier (id) and corresponding prediction values as
obtained by the corresponding model/method. These test identifiers (such as JVASP-1408)  must match the test set ids in the
json.zip file in the benchmarks folder for the metric measurements to work.}}
\end{figure*}

At the homepage, information regarding the number of methods, benchmarks, contributions, and datapoints are provided. A snapshot of the homepage with various categories is shown in Fig. \ref{fig:snapshot}a. Clicking on one of the entries (or searching in the 'Search' box) such as
``formation\_energy\_peratom'' opens a new tab with available contributions. This new tab consists of 1) a description of the benchmark, 2) a plot of various available contributions (as shown in Fig. \ref{fig:snapshot}b), 3) explicit table for the plot (as shown in Fig. \ref{fig:snapshot}c). For each contribution, links
are provided to the submitted data (in .csv.zip format), reference benchmark data (in JSON file), a shell script to reproduce the
contribution (run.sh file) and metadata file (metadata.json). The metadata file contains details about the team name, electronic mail address of the contributor(s), DOI number,
software (with software version), hardware, instrument, computational timing and other relevant details of a benchmark. 

There are several categories for the
benchmarks including AI, ES, QC, FF and EXP and their combinations. Some example contributions and
a summary table are also provided on the webpage to help a user navigate through the project. The summary table breaks down the
available information into categories and sub-categories of different methodologies. 

JARVIS-Leaderboard is an evolving project, so additions to the project are anticipated, welcome, and easy to make. We show a general flowchart for
adding a new benchmark to the leaderboard in Fig. \ref{fig:flowchart}. The user can populate the reference dataset (with well-defined data splits) used
for a specific benchmark (e.g. for 2D exfoliation energies in JARVIS-DFT dataset using an AI method:
``AI-SinglePropertyPrediction-exfoliation\_energy-dft\_3d-test''). AI benchmarks have pre-defined training/validation/test
identifiers and target data in a corresponding json.zip file, while
other methods have only reference test set for evaluation because they do not require model training like an AI method does. For
most benchmarks in the leaderboard, experimental data is used as the reference data. 


There is a helper script jarvis\_populate\_data.py to generate a benchmark dataset. A user can apply their method, train
models, or run experiments on that dataset and prepare a csv.zip file, a metadata.json file, and also if possible, a conda environment.yaml/Nix/Dockerfile and
a run.sh file. This step helps to reproduce the benchmark. These files are kept in a folder with the name of the folder as the team
name and can be uploaded to a user's GitHub account by the automated jarvis\_upload.py script. This script automatically forks the
parent usnistgov/jarvis\_leaderboard repo for the user, adds the team-name folder with its files in that forked repo, runs a few
minimal sanity checks on the new contribution, and then makes a pull request to the parent repo. The contribution addition and
automated testings are carried out using GitHub actions. The administrators of the JARVIS-Leaderboard at NIST will verify the contributions and then finally, it will become part
of the leaderboard website. 

This project is available on GitHub at: \url{https://github.com/usnistgov/jarvis_leaderboard}. The administrators of the JARVIS-Leaderboard at NIST will fully oversee the upload of contributions and benchmarks. A tree structure of the repo is shown
in Fig. \ref{fig:tree}. There are two main directories in the repo: (1) benchmarks (reference) and (2) leaderboard
contributions (for various leaderboard entries), as shown by the green highlighted boxes in Fig. \ref{fig:tree}. 

The ``benchmarks'' directory has folders for the AI, ES, QC, FF, and EXP categories. Within them, there are sub-folders for
specific sub-categories such as (1) SinglePropertyPrediction (where the output of a model/experiment is one single number for an
entry), (2) SinglePropertyClass (where the output is class-ids, i.e., 0,1,.. instead of floating values), (3) ImageClass (for
multi-class image classification), (4) TextClass (for multi-label text classification), (5) MLFF (machine learning force-field), (6)
Spectra (for multi-value data) and (7) EigenSolver (for Hamiltonian simulation). In each of these sub-folders, there are .json.zip
files with well-defined reference datasets and available properties as also available in the JARVIS-Tools package
\url{https://jarvis-tools.readthedocs.io/en/master/databases.html}. To avoid storage of large files in the GitHub repo, the actual
datasets are part of JARVIS-Tools and are stored in the Figshare repository with specific DOIs and version numbers. 

Next, in the ``contributions'' directory, there is a collection of folders that consist of .csv.zip, metadata.json files, and
optionally a Dockerfile and run.sh file. The csv.zip file contains identifier (id) entries and corresponding prediction values
obtained by the corresponding model/method. These test identifiers (such as JVASP-1408 in Fig. (3)) must match the test set IDs in the
json.zip file in the benchmarks folder for the metric measurements to work. Each of the csv.zip files must contain six components in
the filename to place the contribution in the appropriate webpage. The components are the categories (such as AI),
sub-categories (such as ImageClass), property (such as bravais\_lattice), dataset-name (such as stem\_2d\_image as available in the
JARVIS-Tools database page), and data-split. For entries in the AI category, the data is in train-validation-test splits (using a fixed random number generator). For the current leaderboard format, we report the performance accuracy in the test set only. These files can be easily edited with common text editors. Each contribution folder (e.g. alignn-model) consists of one or several csv.zip files corresponding to each benchmark
(such as for formation energies, bandgap, etc.). 

Model-specific details are kept in the metadata.json file with \textit{required}
keys such as model\_name, project\_url, team\_name and an email address. Users can keep other data such as the uncertainty, time taken,
and instrument/software/hardware used in the metadata file as well. For computational models, the run.sh script can be used to reproduce
the contributions completely as a single command line script or job submission script. If a method requires additional
steps or details beyond a simple command line script, a user can upload a README file containing the additional details. For
enhanced reproducibility, we also optionally allow users to include a Dockerfile and an ipython/Google-colab notebook for each
benchmark. These notebooks can be used to run the contributions in the Google-cloud without downloading anything locally.

In addition, there is a ``docs'' directory in the JARVIS-leaderboard. The docs folder consists of a directory structure that is
similar to the benchmarks folder with categories names (AI, ES, etc.), and sub-categories (such as SinglePropertyPrediction,
ImageClass etc.) with markdown (.md) files that will be converted automatically into corresponding html pages for the website.
For each benchmark (i.e., json.zip file), a corresponding docs entry (i.e., md file) should be present. 
A
new benchmark must be associated with a peer-reviewed article and a DOI, in order to have trust in the reference benchmark
data. A new benchmark must also be verified by the JARVIS-Leaderboard administrators. 

 \begin{table}[!hbt]
\caption{Comparison of benchmark infrastructure available for materials design methods for several categories. }\label{tab:classification}
    \centering
    \begin{tabular}{|l|l|l|l|l|l|l|l|}
    \hline
        Projects & AI& ES& FF& QC& EXP\\ \hline
        MoleculeNet\cite{wu2018moleculenet} & $\checkmark$ &-&-&-&-\\ \hline
        MatBench\cite{dunn2020benchmarking} & $\checkmark$ &-&-&-&-\\ \hline
        OpenCatalystProject\cite{chanussot2021open} & $\checkmark$ &-&-&-&-\\ \hline
        
        SciML\cite{scimlbench:2021} & $\checkmark$ &-&-&-&-\\ \hline
        SGDML\cite{chmiela2019sgdml} & $\checkmark$ &-&-&-&-\\ \hline
        GuacaMol\cite{brown2019guacamol} & $\checkmark$ &-&-&-&-\\ \hline
        Alchemy\cite{chen2019alchemy} & $\checkmark$ &-&-&-&-\\ \hline
        ML4Chem\cite{khatib2020ml4chem} & $\checkmark$ &-&-&-&-\\ \hline
        DGL-LifeSci\cite{broccatelli2022benchmarking} & $\checkmark$ &-&-&-&-\\ \hline
        CCCBDB\cite{johnson2006nist}&-&$\checkmark$&-&-&$\checkmark$\\ \hline 
        Delta-DFT\cite{lejaeghere2016reproducibility}&-&$\checkmark$&-&-&-\\ \hline  
        SSSP\cite{prandini2018precision}&-&$\checkmark$&-&-&-\\ \hline 
        OpenKIM\cite{karls2020openkim}&-&-&$\checkmark$&-&-\\ \hline 
        IPR\cite{hale2018evaluating}&-&-&$\checkmark$&-&-\\ \hline 
        JARVIS-FF\cite{choudhary2018high}&-&-&$\checkmark$&-&-\\ \hline 
        Mlearn\cite{zuo2020performance}&-&-&$\checkmark$&-&-\\ \hline 
        QuantumVolume\cite{cross2019validating}&-&-&-&$\checkmark$&-\\ \hline
        SupermarQ\cite{tomesh2022supermarq}&-&-&-&$\checkmark$&-\\ \hline
        Olympus\cite{olympus}&-&-&-&-&$\checkmark$\\ \hline
        Golem\cite{aldeghi2021golem}&-&-&-&-&$\checkmark$\\ \hline
        HTE-MC\cite{hattrick2019inter}&-&-&-&-&$\checkmark$\\ \hline
        JARVIS-LB& $\checkmark$ & $\checkmark$ & $\checkmark$ & $\checkmark$ & $\checkmark$\\ \hline

    \end{tabular}
\end{table}

As mentioned above, there already exist several other materials science-specific benchmarks. We compare some of these benchmarks in
Table \ref{tab:classification} based on the categories that are included. We find that there is no single, large-scale benchmark
encompassing the various fields as in the JARVIS-Leaderboard. Also, the data format, metadata, and website for these different
leaderboards vary significantly. Hence, having a uniform way to compare different methods would greatly help the materials
community.




\subsection{Benchmarks}
\begin{figure*}[hbt!]
    \centering
    \includegraphics[trim={0. 0cm 0 0cm},clip,width=0.98\textwidth]{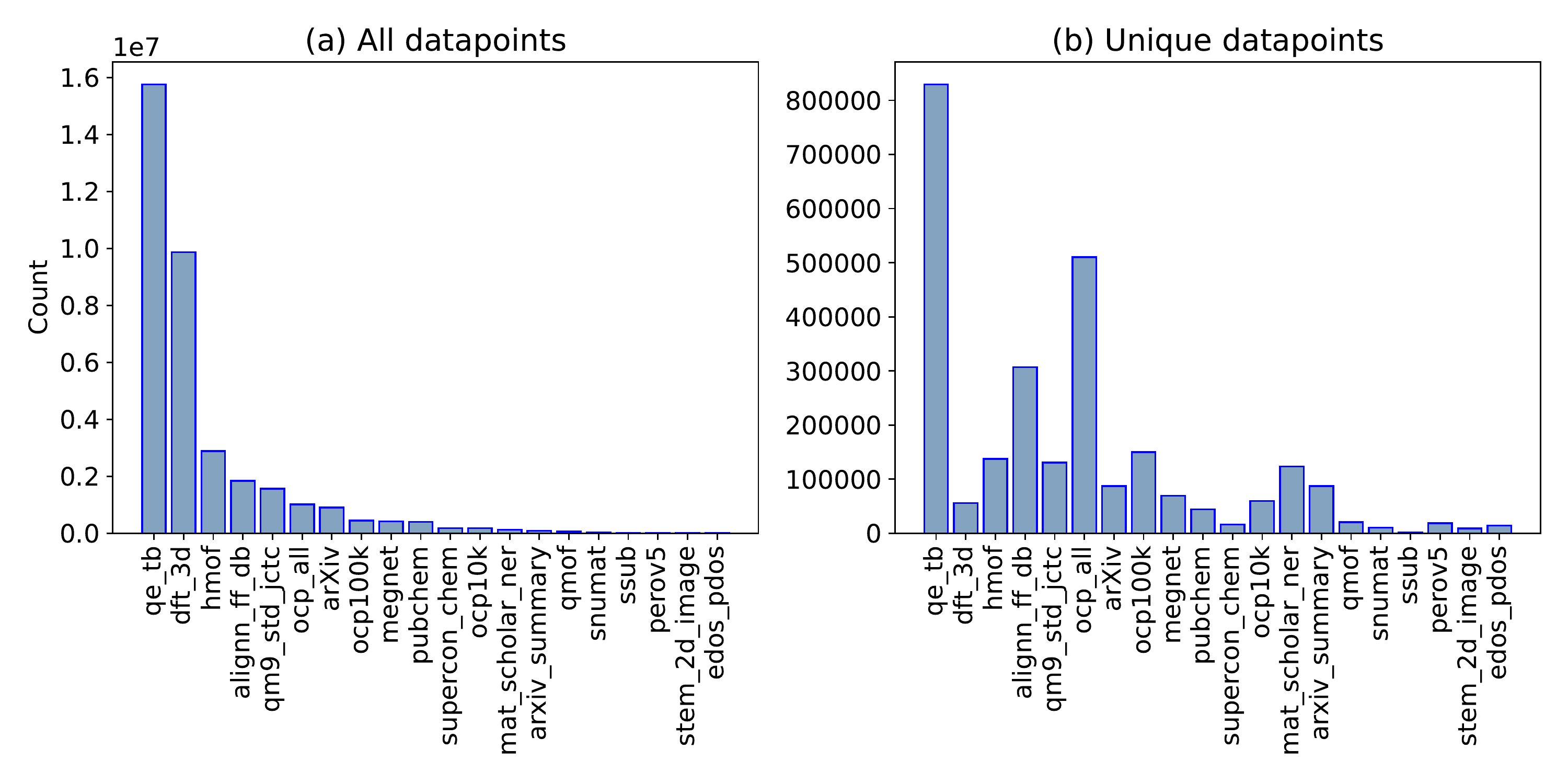}
    \caption{\label{fig:dataset}{Distribution of data in each dataset. (a) all entries in leaderboard, (b) entries with unique identifiers. Note that one identifier (such as JVASP-1002 for silicon) can have multiple properties (such as bandgap, bulk modulus etc.). A script to generate this figure is also provided on the leaderboard website as the leaderboard is continuously evolving.}}
\end{figure*}


\begin{figure*}[hbt!]
    \centering
    \includegraphics[trim={0. 0cm 0 0cm},clip,width=0.98\textwidth]{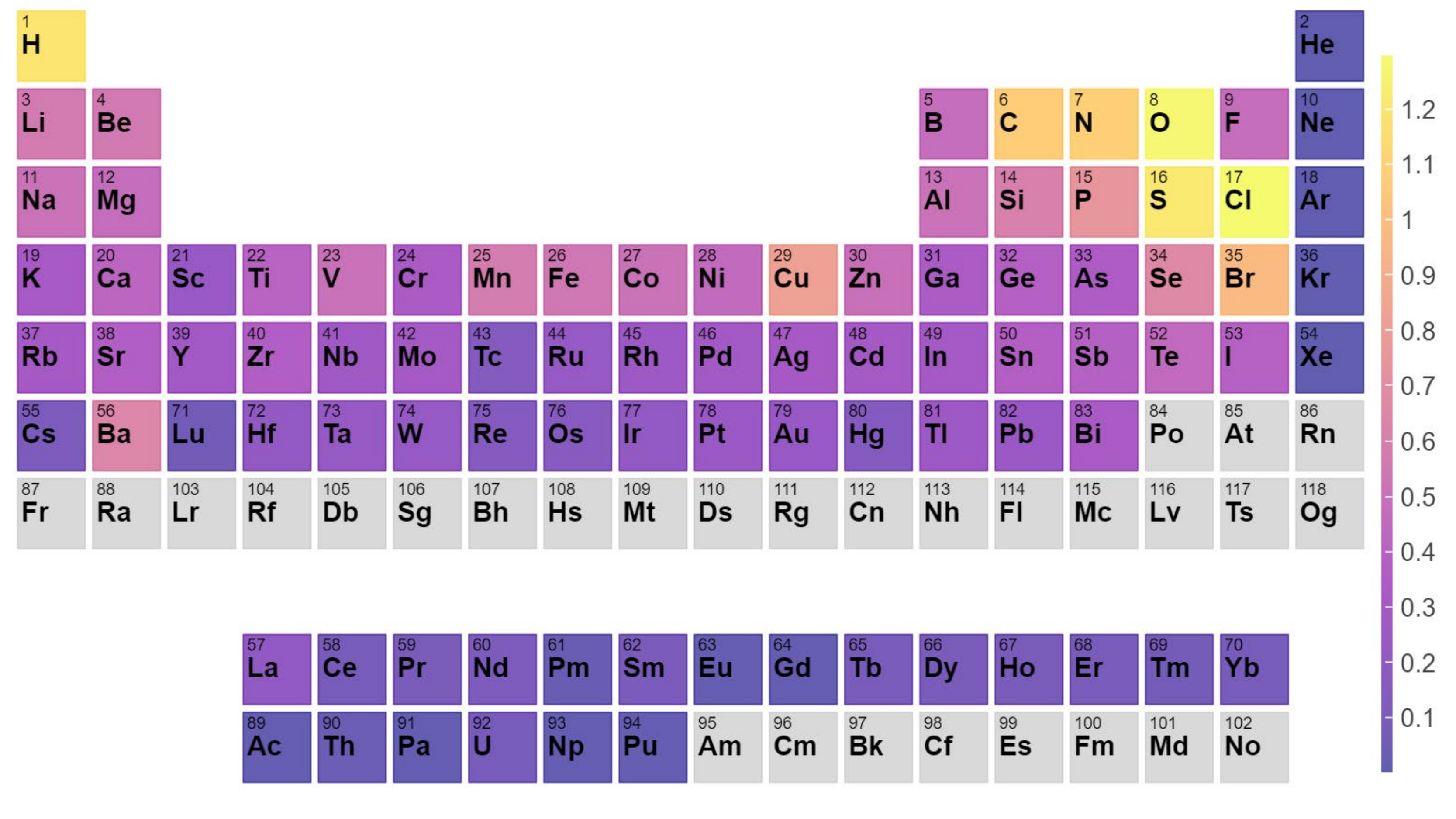}
    \caption{\label{fig:periodic}{Periodic table element distribution for entries in all the datasets. This is calculated by taking into account all the element specific entries normalized by total entries i.e. these are percentage probabilities.}}
\end{figure*}

The benchmarks consist of experimental data, density functional theory, or numerical solutions that are well-known and have already been published in peer-reviewed articles or books. A benchmark should be considered the ``ground truth" for a particular task. Therefore, it is mandatory to have a digital object identifier (DOI) for each benchmark from a peer-reviewed article. There can be multiple contributions from different models or experiments for a benchmark, e.g., contributions from various DFT functionals in predicting the electronic bandgap of silicon with respect to experimental data. Typically, for electronic structure (ES) method based contributions, the benchmarks are experimental data; for artificial intelligence (AI) methods, they are the test split; for force-field (FF) methods, they are electronic structure data; for quantum computation (QC), they are analytical results; and for experiments (EXP), they are other experiments. Currently, we have more than
270 benchmarks in the leaderboard. The JARVIS-Leaderboard flexible and dynamic nature allows addition of new benchmarks as well. 

Each entry in the benchmark dataset consists of a unique identifier. Most of these datasets are integrated
into JARVIS-Tools databases page already (but not limited by it), with an associated JARVIS ID number (JID) and are backed up in Figshare, Google Drive and NIST-internal
storage systems. The number of entries can vary from a few (which is especially applicable for experimental and high-accuracy computational methods,
where generating a very large dataset is not feasible in terms of time and resources) to hundreds of thousands of entries in a dataset.

An overview of the dataset can be found in Fig. \ref{fig:dataset}. Considering all possible entries in the dataset, we have close to
7 million datapoints. For example, an atomic structure can have multiple properties calculated, such as bandgaps and formation
energies, among other properties. We find the JARVIS-DFT-3D dataset to have the largest number of entries. Considering unique systems, we can find the distribution in Fig. \ref{fig:dataset}b). In this case, qe-tb (fitting dataset for ThreeBodyTB.jl  \cite{PhysRevMaterials.7.044603}) is one
of the largest datasets available in the leaderboard. Note that these datasets contain all varieties of data modalities such as
atomic structure, images, spectra and text. In Fig. \ref{fig:periodic}, we show the fractional distribution of periodic table elements in the entire
dataset. We find that the most common elements are C, N, O, Cu which is similar to the natural abundance of these elements.

Experimental results are uploaded as benchmarks (i.e. what is regarded as the reference). In the absence of
experimental data, high-fidelity computational methods can be used as a reference. If there are multiple experimental measurements available in the literature, each can be individually added as separate benchmarks (i.e., different json.zip files to distinguish one benchmark from another) and users can submit contributions for each of them. As time and the materials science field progresses, certain experimental data may need to be revisited (i.e. more accurate measurements in the future or results are reported that contradict previous experimental data). As a response to this, separate reference (experimental) benchmarks can be added, and users will be able to plot and compare the evolution of these benchmarks over time. 

In addition, Leaderboard users can raise an issue on GitHub pertaining to reference benchmarks. The administrators will also upload a README file which contains additional information about the experiments conducted, including associated DOI, experimental conditions and provide details if additional experiments conducted on the same material/property exist in the literature. The experimental conditions described in the README file can be important when comparing the reference benchmark to calculated results, which may be in different conditions than the experiment (i.e. the bandgap of a material is never measured at 0 K, as DFT predicts).

Contributions to the leaderboard in the
form of user-submitted experimental data can be compared with previous experiments, electronic structure methods or other numerical
results.  ES-based contributions are benchmarked against experimental results and can be compared with other ES methods. 
QC data can be compared with classical computation data or exact analytical results. For FF, contributions can be compared to DFT (or other ES data) or high-level interatomic potential benchmark suites (specifically for MLFFs) \cite{mlff}. For AI, a test dataset is used.
Unlike other methods, AI methods can have both ``train'' and ``test'' datasets, while others have only ``test'' sets in the corresponding
dataset. For AI methods, if the ``train'' dataset is not provided and only ``test'' is given, the benchmark can be used for checking
extrapolation behavior such as vacancy formation energy benchmarks.



\subsection{Analysis of Benchmarks}

Presently, the leaderboard has 5 categories, 10 sub-categories, 152 methods,  274 benchmarks, 1281 contributions and 8714228
datapoints. In this section, we show a few of the hundreds of example analyses that can be carried out using the available
benchmarks and contributions. In Fig. \ref{fig:mae_example}, we show the MAE of the AI computed formation energy and ES computed
bandgap for Si for a variety of contributions in the leaderboard. In Fig. \ref{fig:mae_example}(a) we see the comparison of 12 AI
models (each AI model had a well-defined 80:10:10 split for training, validation and testing respectively from the JARVIS-3D database) and find the kgcnn\_coGN \cite{REISER2021100095} has the highest accuracy/lowest error, followed by Potnet\cite{lin2023efficient}, Matformer\cite{yan2022periodic} and ALIGNN\cite{Choudhary2021_ALIGNN, gupta2024structure} models. This can be attributed to
the fact that as we include more structural information and use deep-learning methods rather than descriptor methods, we get improvement
in accuracy. 

Similarly, in Fig. \ref{fig:mae_example}(b) we compare the bandgap of Si using several methods and find GLLB-sc \cite{PhysRevB.82.115106} calculated with GPAW \cite{Enkovaara_2010} to yield
the lowest error, while G$_0$W$_0$ \cite{RevModPhys.74.601} (VASP \cite{kresse1996efficient,kresse1996efficiency}), GW$_0$ \cite{RevModPhys.74.601} (VASP \cite{kresse1996efficient,kresse1996efficiency}), TBmBJ \cite{mbj,mbj-2} (VASP), and DMC \cite{RevModPhys.73.33} (QMCPACK \cite{kim2018qmcpack}) methods follow. This can be attributed to the inclusion of the discontinuity potential (GLLB-sc \cite{PhysRevB.82.115106}) or
kinetic energy density (TBmBJ \cite{mbj,mbj-2}) in the density functional or incorporating many-body physics (G$_0$W$_0$ \cite{RevModPhys.74.601}, GW$_0$ \cite{RevModPhys.74.601}, DMC \cite{RevModPhys.73.33}) into the methodology, which can lead to improved accuracy for bandgap prediction. Also, similar methods such as PBE\cite{PhysRevLett.77.3865}
data from Open Quantum Materials Database (OQMD) \cite{oqmd,oqmd-2}, AFLOW \cite{CURTAROLO2012218} and Materials Project \cite{doi:10.1063/1.4812323} compare well with each other. 

\textcolor{black}{In Fig. \ref{fig:mae_example}(c) we compare how several classical FFs compute the Voigt bulk modulus of Si. In Fig \ref{fig:mae_example}(d) we compare several MLFF models for the forces of Si. We compare various pretrained MLFFs and other MLFFs we specifically trained on the MLEARN \cite{zuo2020performance} dataset (PBE-based DFT data). We see that ALIGNN-FF \cite{choudhary2023unified} and MatGL \cite{m3gnet} perform similarly for prediction of forces. Fig. \ref{fig:mae_example}(d) provides a comprehensive comparison of MLFFs that are trained and tested on the same dataset and pretrained models that were trained elsewhere.} The comparisons are presented in tabular form for all the benchmarks on
the leaderboard website. We have provided tools and notebooks in the leaderboard GitHub repository that can be used for making such
plots for all the available benchmarks and contributions. A collection of such figures for method comparison is available in the supplementary information ( Supplementary Figures 1-298).  We have also added interactive plots for such comparisons on the
website. These tools can aid in identifying examples of materials that require high-fidelity methods beyond the accuracy of DFT in order to understand their underlying properties. In addition, these tools can be used to validate electronic structure methods and provide insight for error estimation. 

The leaderboard has a large number of benchmarks and can enable a more comprehensive comparison of different methods for better revealing their respective advantages and limitations. For instance, neural networks outperform descriptor-based models by a large degree in all of the 10 regression tasks in the latest Matbench~\cite{dunn2020benchmarking} leaderboard. To check if this is also the case for 44 regression benchmarks in the current JARVIS leaderboard, we compare the performance of the best descriptor-based model to that of the best neural network. As shown in Fig.~\ref{fig:matminer}, the best neural network outperform the best descriptor-based model in 34 tasks, but only 14 out of 44 (32 \%) tasks see a performance difference by more than 20 \%. This indicates that descriptor-based models are still competitive with respect to neural networks, especially considering their better interpretability and orders of magnitude lower training cost~\cite{li2023critical,li2023redundancy}. Notably, the best descriptor-based model is found to outperform the best neural network in 10 tasks including those with $10^4$-$10^5$ training data, opening up interesting questions and potential direction to further model improvement. For instance, the inferior performance of neural networks in the regression tasks for the heat capacity and hMOF data may be related to the recently revealed incapability of graph neural networks in capturing periodicity~\cite{gong2022examining}.

\subsection{Analysis of Error Metrics}
Although a metric such as the MAE can be useful to compare methods for a specific benchmark, it is difficult to compare across
different methods, since MAE values can differ substantially. Hence, we use the mean absolute deviation (MAD, computed with respect to the average value of the training data as a baseline/random-guess model) to MAE ratio for both AI and
ES single-property-prediction categories. Mean absolute deviation values act as a baseline/random-guessing model for the benchmark
and contributed models should have MAE performance better than MAD values. We show the MAD/MAE ratios for AI and ES benchmarks in
Fig. \ref{fig:madmae}. We find that the MAD/MAE values range from 2 to 50. MAD/MAE values close to 1 suggest low predictive power.
We observe that quantum properties such as the bandgap have lower MAD/MAE than classical quantities (quantities that do not require quantum mechanical simulations) such as total energy or bulk
modulus. Interestingly, such trends for classical vs. quantum quantities are observed for both the AI and ES approaches. 

\subsection{Interactive View of Benchmarks and Contributions}
\begin{figure*}[hbt!]
    \centering
    \includegraphics[trim={0. 0cm 0 0cm},clip,width=0.98\textwidth]{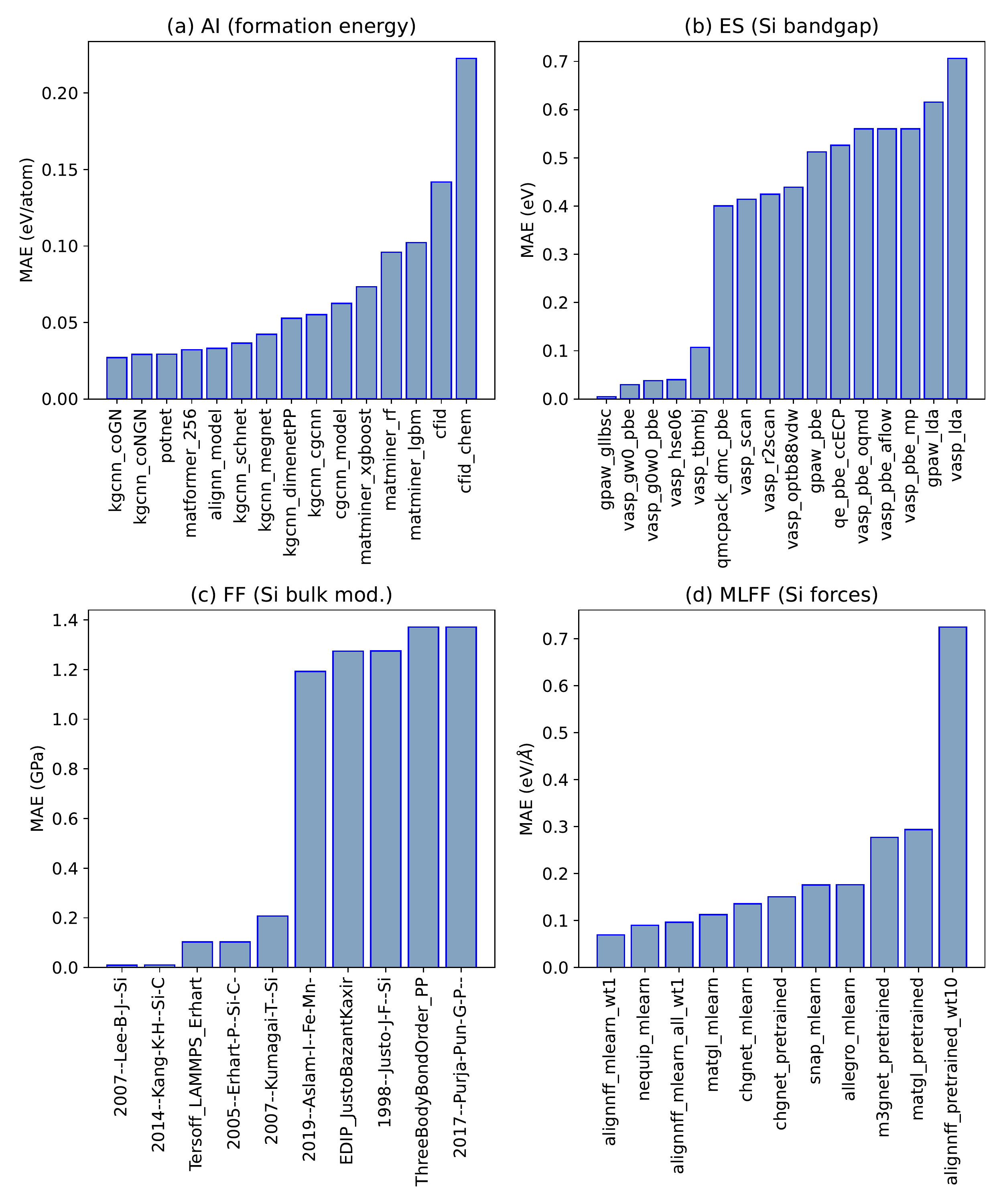}
    \caption{\label{fig:mae_example}{Example mean absolute errors for benchmarks including (a) artificial intelligence (AI) formation
energy for test set with 5572 materials in JARVIS-DFT 3D dataset, (b) electronic structure (ES) Si (JARVIS-DFT ID: JVASP-1002) bandgap, \textcolor{black}{(c) classical force-field (FF) based Voigt bulk modulus of Si and (d) machine learning force-field (MLFF) based forces for Si. We provide Jupyter/Google colab notebooks to easily plot such comparisons for all available benchmarks. Also, similar analysis figures for all the available benchmarks are available in the supplementary information (Supplementary Figures 1-298). As a note, these plots are a current snapshot of the leaderboard, and it is possible that new and more accurate models will be developed and added here in the future.}}}
\end{figure*}

\begin{figure*}[!hbt]
    \centering
    \includegraphics[trim={0. 0cm 0 0cm},clip,width=\textwidth]{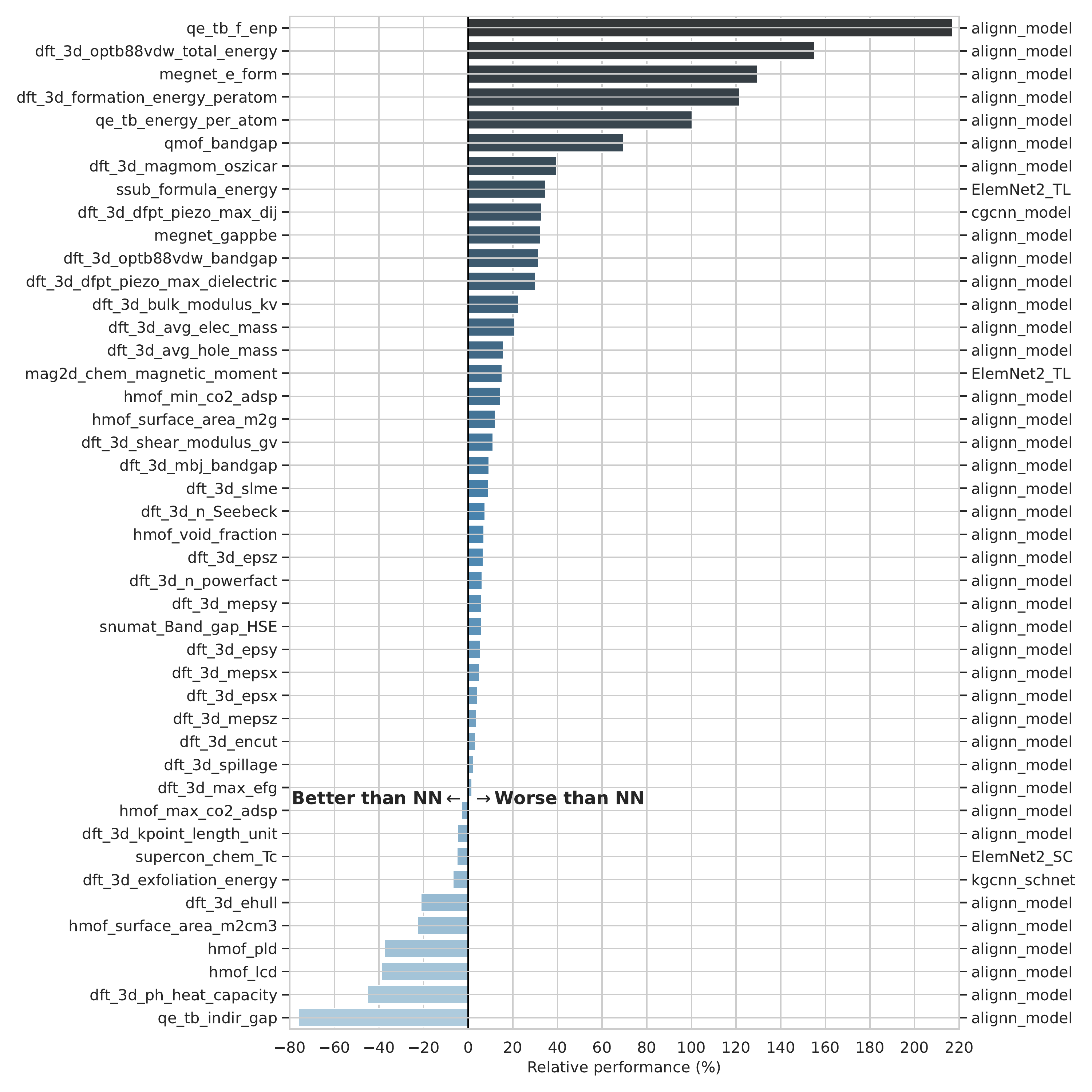}
    \caption{\label{fig:matminer}{Relative performance computed as the ratio of the MAE difference between the best descriptor-based model and the best neural networks to the MAE of the best neural networks in the AI regression benchmarks. The benchmark name and the corresponding best performing neural network are indicated in the left and right y axis, respectively. For all the considered AI benchmarks, the best descriptor-based model is the tree-based model using Magpie \cite{magpie} and Voronoi-tessellation \cite{ward2017including} features. \textcolor{black}{As a disclaimer, these plots are a current snapshot of the leaderboard, and it is possible that new and more accurate models will be developed in the future.} }}
\end{figure*}

\begin{figure*}[ht!]
    \centering
    \includegraphics[trim={0. 0cm 0 0cm},clip,width=0.98\textwidth]{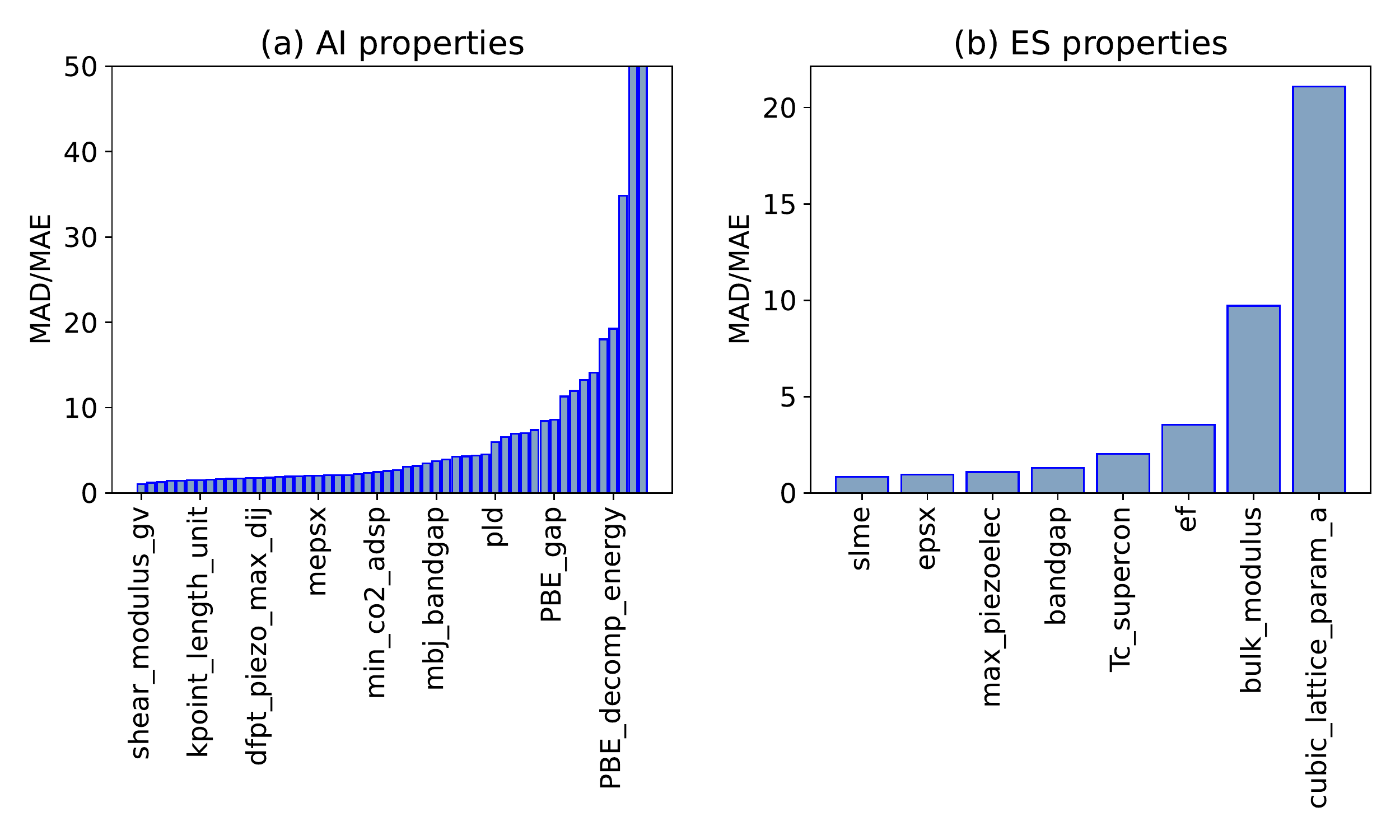}
    \caption{\label{fig:madmae}{Mean absolute deviation (MAD) to mean absolute error (MAE) ratio for (a) AI and (b) electronic structure methods. MAD:MAE serves as uniform criteria for comparing performances of models.}}
\end{figure*}

\begin{figure*}[hbt!]
    \centering
    \includegraphics[trim={0. 0cm 0 0cm},clip,width=0.98\textwidth]{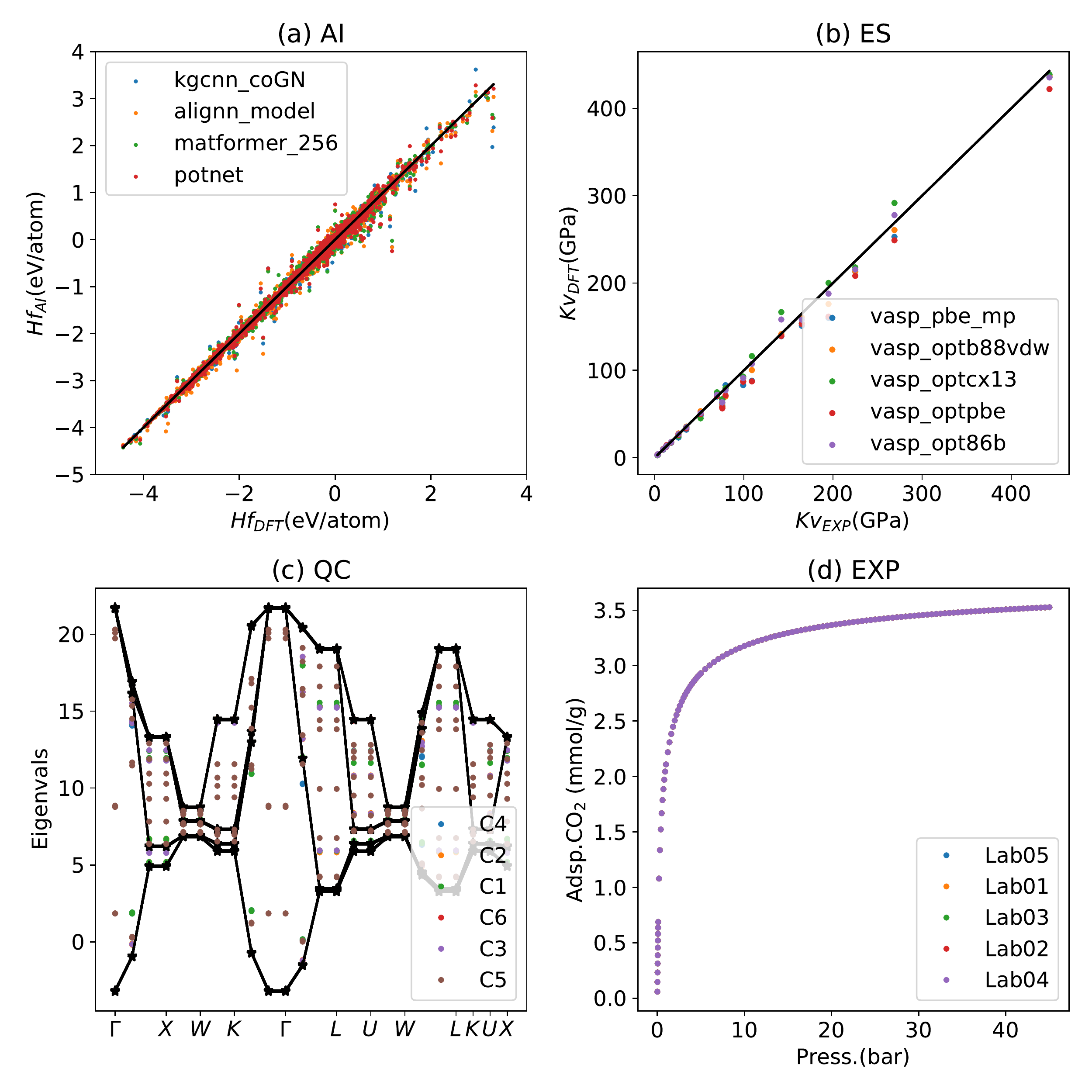}
    \caption{\label{fig:exresults}{Example results for AI, ES, QC and EXP results. (a) formation-energy-per atom model using AI for JARVIS-DFT 3D dataset with 5572 materials in the test set, (b) bulk modulus predictions using ES methods for 21 materials, (c) electronic bandstructure of Aluminum using QC methods with different quantum circuits on a coarse k-point mesh, (d) CO$_2$ capture for zeolite (ZSM-5) at several labs in inter-laboratory/round-robin fashion.}}
\end{figure*}

In addition to making bar plots as shown in Fig. \ref{fig:mae_example} and Fig. \ref{fig:madmae}, the raw data available in
benchmarks and contributions can be presented in various other forms such as scatter plots, bandstructures, adsorption spectra, and
diffraction spectra. In Fig. \ref{fig:exresults}, we show example comparisons of different methods for AI, ES, QC and EXP categories
including (a) formation-energy-per atom model using AI, (b) bulk modulus predictions using ES, (c) electronic bandstructure of Al using
QE with different quantum circuits \cite{ChoudharyAtomQC}, (d) CO$_2$ capture for zeolite at several labs in round-robin fashion \cite{co2-rr}. In Fig.
\ref{fig:exresults}a), we find that formation energy is one of the easiest quantities to train AI models and even simple chemistry
only-based models can perform reasonably well (i.e., cfid\_chem). Including more structural features (such as bond angles and dihedral angles) and using deep learning models (such as graph neural network vs descriptor based models)
further helps improve accuracy. Similarly, for ES example for predicting bulk modulus, we find irrespective of DFT based method
used, they are in relatively close agreement with experimental bulk modulus data as shown in Fig. \ref{fig:exresults}b). In Fig
\ref{fig:exresults}c), we find that the selection of a quantum circuit is critically important for predicting electronic band
structures well. Here, we used 6 different quantum \cite{ChoudharyAtomQC} circuits and found the SU(2) \cite{Qiskit} circuit to compare well with classical computer-based
electronic bandstructures. This can be attributed to various entanglements captured in the SU(2) \cite{Qiskit} circuits that may be missing in other
circuits. Finally, for experimental inter-laboratory/round-robin type measurements of the zeolite CO$_2$ isotherm, we find excellent agreement across
different labs \cite{co2-rr}.


\section{Methods}
The JARVIS-Leaderboard aims to provide a comprehensive framework covering a variety of length and time-scale approaches
\cite{callister2000fundamentals} to enable realistic materials design. In this section, we provide a brief overview of the methods
that are currently available in the leaderboard. In this work we use the terms categories, sub-categories, methods, benchmarks, and
contributions often, so we define them as follows.

 Currently, there are five main ``categories'' in the leaderboard: Artificial Intelligence (AI), Electronic Structure (ES),
 Force-field (FF), Quantum Computation (QC), and Experiments (EXP). Each category is divided into ``sub-categories", a
 list of which is provided on the website. These sub-categories include single-property-prediction, single-property-classification,
 atomic force prediction, text classification, text-token classification, text generation, image classification,
 image segmentation, image generation, spectra-prediction, and eigensolver. These sub-categories are highly flexible and new
 categories can be easily added. ``Benchmarks" are the reference data (in the form of json.zip file, discussed later) used to
 calculate performance metrics for each specific contribution. ``Methods" are a set of precise specifications for evaluation against
 a benchmark. For example, within the ES category, density functional theory (DFT) performed with the specifications of the Vienna
 Ab initio Simulation Package (VASP)\cite{kresse1996efficient,kresse1996efficiency},  Perdew-Burke-Ernzerhof (PBE)
 \cite{PhysRevLett.77.3865} functional and PAW \cite{kresse1996efficient,kresse1996efficiency} pseudopotentials (VASP-PBE-PAW) is a
 method. Similarly, within the AI category, descriptor/feature-based models with specifications of MatMiner \cite{WARD201860} chemical features and
 the LightGBM \cite{ke2017lightgbm}software is a method. ``Contributions" are individual data (in the form of csv.zip files) for each benchmark computed with
 a specific method. Each contribution files consist of six components: category (e.g. AI), sub-category (e.g.
 SinglePropertyPrediction), property (e.g. formation energy), dataset (e.g. dft\_3d), data-split (e.g. test), metric (e.g. mae).

\subsection{Electronic structure}
Electronic structure approaches cover short length scales and short time scales with high-fidelity. There are a variety of ES
methodologies such as such as tight-binding \cite{harrison2012electronic,garrity2021database,PhysRevMaterials.7.044603}, density
functional theory (DFT)\cite{martin2020electronic}, quantum Monte Carlo \cite{RevModPhys.73.33}, dynamical mean field theory
\cite{kotliar2006electronic} and many-body perturbation theory (Green's function with screened Coulomb potential, GW
methods)\cite{RevModPhys.74.601}. For each of the methodologies, there are a number of specifications to completely describe a method
including the exact software, exchange-correlation functional, pseudopotential, and other relevant parameters. Example methods used
in this work are given in Table \ref{tab:methods}.

\begin{table*}[]
\caption{Summary of current benchmark categories and methods available in the JARVIS-Leaderboard at the time of writing. More details can be found in the individual metadata.json file. Note that the number of methods is continuously growing.}\label{tab:methods}
\begin{tabular}{|l|l|l|}
\hline
Category & General name & Method Specification                            \\
\hline
ES       & DFT \cite{PhysRev.136.B864}    & VASP\cite{kresse1996efficient,kresse1996efficiency} (PBE\cite{PhysRevLett.77.3865}, LDA \cite{PhysRev.136.B864}, OptB88vdW \cite{klimevs2009chemical}, \\
& & Opt86BvdW \cite{PhysRevB.83.195131}, TBmBJ \cite{mbj,mbj-2}, SCAN \cite{PhysRevLett.115.036402}, r2SCAN \cite{r2scan}, HSE06 \cite{doi:10.1063/1.1564060}) \\\cline{3-3}

         &        & QE \cite{giannozzi2009quantum} (PBE\cite{PhysRevLett.77.3865}, PBEsol \cite{PhysRevLett.100.136406})                                \\\cline{3-3}
         &        & ABINIT \cite{gonze2016recent, Romero2020,Gonze2020} (PBE\cite{PhysRevLett.77.3865})                                \\\cline{3-3}         
         &        & GPAW \cite{Enkovaara_2010} (PBE\cite{PhysRevLett.77.3865}, LDA \cite{PhysRev.136.B864}, GLLB-sc \cite{PhysRevB.82.115106})                                \\\cline{2-3}         
         & QMC\cite{RevModPhys.73.33}    & QMCPACK\cite{kim2018qmcpack} (DMC\cite{RevModPhys.73.33})             \\\cline{2-3}
        & GW\cite{RevModPhys.74.601}    & VASP\cite{kresse1996efficient,kresse1996efficiency} (G$_0$W$_0$ \cite{RevModPhys.74.601}, GW$_0$ \cite{RevModPhys.74.601}) 
        \\\cline{2-3}
         & TB \cite{harrison2012electronic}     & ThreeBodyTB.jl  \cite{PhysRevMaterials.7.044603} (Wannier90 \cite{MOSTOFI20142309}) \\      
\hline  

AI       & Descriptor    & CFID \cite{choudhary2018machine}, MagPie \cite{magpie}, MatMiner \cite{WARD201860,li2023redundancy}, crystal feature model \cite{wexler2021p13212}, \\
& & ElemNet \cite{elemnet, jha2019enhancing, elemnettl, gupta2023pre}, IRNet\cite{jha2019irnet, irnet, irnettl}, BRNet\cite{brnet, branchnet, ibrnet}, SNAP \cite{PhysRevMaterials.1.043603}\\\cline{2-3}
         & Graph-based    & ALIGNN \cite{Choudhary2021_ALIGNN, gupta2024structure}, CGCNN \cite{Xie2018CGCNN}, SchNet \cite{Schutt2018SchNet}, AtomVision \cite{Choudhary2023_AtomVision}, \\
         & &  ChemNLP \cite{ChemNLP_arxiv}, DimeNet+ \cite{gasteiger_dimenet_2020,gasteiger_dimenetpp_2020}, CHGNet \cite{deng2023_chgnet}, M3GNET \cite{m3gnet}\\,
         & & kgcnn\_coGN \cite{REISER2021100095}, Potnet\cite{lin2023efficient}, Matformer\cite{yan2022periodic}\\\cline{2-3}
        & Transformers    & OPT \cite{zhang2022opt}, GPT \cite{brown2020language}, T5 \cite{raffel2020exploring}\\\cline{2-3}

\hline  

FF       & LJ \cite{doi:10.1098/rspa.1924.0081}   & LAMMPS \cite{LAMMPS} (2D-Liquid)\\\cline{2-3} 
         & EAM \cite{PhysRevB.29.6443}  & LAMMPS \cite{LAMMPS} (FCC-Al) \\\cline{2-3}
         & REBO \cite{PhysRevB.37.6991}   & LAMMPS \cite{LAMMPS} (Diamond-Si) \\\cline{2-3}
        & AMBER99sb-ildn  \cite{lindorff2010improved}  & GROMACS \cite{van2005gromacs} (Alanine dipeptide) \\\cline{2-3}         
        & CHARMM36m  \cite{mehdi2022accelerating}  & GROMACS \cite{van2005gromacs} ($\alpha$-aminoisobutyric acid) \\\cline{2-3}

\hline 

QC       & Algorithms    & Qiskit \cite{Qiskit} (VQE \cite{vqe}, VQD \cite{Higgott2019variationalquantum})\\\cline{3-3} 
         &    & PennyLane \cite{bergholm2022pennylane,arrazola2023differentiable} (VQE \cite{vqe}, VQD \cite{Higgott2019variationalquantum}) \\\cline{2-3}
         & Circuits   & Qiskit \cite{Qiskit} (PauliTwo Design \cite{Qiskit}, SU(2) \cite{Qiskit}) \\\cline{2-3}

\hline

EXP       & Diffraction    & XRD (Bruker D8)\\\cline{2-3} 

         & Manometry   & CO$_2$ adsorption FACT lab \cite{co2-rr}\\\cline{2-3}
         & Vibroscopy   & Kevlar FAVIMAT \cite{doi:10.1177/0040517520918232}\\\cline{2-3}
         & Magnetometry   & Susceptibility (PPMS) \cite{doi:10.1177/0040517520918232}\\\cline{2-3}

\hline

\end{tabular}
\end{table*}


Each method in the ES category can have a variety of contributions. For example, using a specific method, one can calculate various
properties such as bandgaps, formation energies, bulk moduli, solar cell efficiencies, and superconducting transition temperatures
as well as spectral quantities such as dielectric functions. While there are more than 400 approximate exchange-correlation
functionals proposed in DFT literature \cite{lehtola2018recent}, currently, we have OptB88vdW \cite{klimevs2009chemical}, Opt86BvdW
\cite{PhysRevB.83.195131}, LDA \cite{PhysRev.136.B864}, PBE\cite{PhysRevLett.77.3865}, PBEsol \cite{PhysRevLett.100.136406}, GLLB-sc \cite{PhysRevB.82.115106}, TBmBJ \cite{mbj,mbj-2}, SCAN
\cite{PhysRevLett.115.036402}, r2SCAN \cite{r2scan}, HSE06 \cite{doi:10.1063/1.1564060}, in the leaderboard. We use converged
k-points and cut-offs as available in the JARVIS-DFT database \cite{choudhary2019convergence}. We have used the Vienna Ab initio Simulation Package (VASP)
\cite{kresse1996efficient,kresse1996efficiency}, ABINIT \cite{gonze2016recent, Romero2020,Gonze2020}, GPAW \cite{Enkovaara_2010} and Quantum Espresso
(QE) \cite{giannozzi2009quantum} as DFT software packages, but other packages can be easily added as well. In addition, we use VASP \cite{kresse1996efficient,kresse1996efficiency}
to perform GW calculations including ``single-shot" G$_0$W$_0$ and self-consistent GW$_0$ methods \cite{RevModPhys.74.601}. Other ES
approaches include tight-binding (TB) \cite{harrison2012electronic} and quantum Monte Carlo (QMC) \cite{RevModPhys.73.33}. For TB,
we use the recently developed ThreeBodyTB.jl code\cite{PhysRevMaterials.7.044603} along with the Wannier90 \cite{MOSTOFI20142309}
code, while the QMCPACK \cite{kim2018qmcpack} code is used for diffusion Monte Carlo (DMC) \cite{RevModPhys.73.33} calculations.

\subsection{Force-field}
Force fields can be used in molecular dynamics and Monte Carlo simulations for studying larger time and length scales compared to electronic
structure methods. Traditional force fields are developed for specific chemical systems and applications and may not be transferable
to other uses. It is important to check the validity of an FF before using it in a particular application. Moreover, the development
of FFs is a cumbersome task. Examples of typical FFs include embedded-atom method (EAM) potentials \cite{PhysRevB.29.6443} ({i.e. Al099.eam.alloy for aluminum system \cite{ff}), Lennard Jones (LJ)\cite{doi:10.1098/rspa.1924.0081} for 2D liquids, reactive empirical bond order (REBO)\cite{PhysRevB.37.6991} for Si, and classical, atomistic force fields for biomolecular systems\cite{Case2005,Huang2017}.
Recently, machine learning force fields (MLFF)\cite{NOVOSELOV201946, PhysRevB.99.014104, PhysRevB.87.184115, PhysRevLett.120.143001, https://doi.org/10.1002/qua.24836, Smith2021} have become popular because of their higher accuracy and ease of development (such as
SNAP \cite{PhysRevMaterials.1.043603} FFs). Nevertheless, early generations of MLFFs were also developed for specific types of chemistry and applications. Very recently,
several MLFFs have been developed that can be used to simulate any combination of periodic table elements. Some of these FFs include
M3GNET \cite{m3gnet}, ALIGNN-FF \cite{choudhary2023unified}, and CHGNet \cite{deng2023_chgnet}. In the leaderboard, we include benchmarks for energies, forces, and stress tensors for both specific
systems and universal datasets.

Traditional FFs are available in LAMMPS \cite{LAMMPS}, while MLFFs are integrated into the Atomic Simulation Environment (ASE) \cite{ase-paper} package. Some of these MLFFs are now available
in LAMMPS and other large-scale MD codes. In addition to static quantities, FFs can be used for Monte Carlo simulations, such as CO$_2$
adsorption in metal-organic frameworks (MOFs) \cite{CHOUDHARY2022111388} using the RASPA \cite{doi:10.1080/08927022.2015.1010082} code. In addition to energy, force, and stress, we also have FF benchmarks for classical
properties such as the bulk modulus. For biomolecular systems, GROMACS\cite{Pall2020} is commonly used, and we present here free energy differences and conformational state population benchmarks for three model peptides\cite{Tsai2021sgoop,Mehdi2022,Wang2021SPIB}.




\subsection{Artificial intelligence}

Recently artificial intelligence methods have become popular for materials prediction across all lengths and time scales. We
currently have benchmarks for four types of data used as input for the AI models: (1) atomic structure, (2) spectra, (3) images, and 4)
text. AI techniques can be used for both forward prediction and  inverse design. For atomic structure datasets, we use DFT datasets such as JARVIS-DFT\cite{choudhary2020joint,10.1063/5.0159299}, Materials Project (MP) \cite{doi:10.1063/1.4812323}, Tight binding three-body dataset (TB3) \cite{PhysRevMaterials.7.044603}, Quantum-Machine 9 (QM9) \cite{qm9-2,ramakrishnan2014quantum}. For spectral data, we use either DFT-based spectra
of, for example, electron or phonon density of states (DOS), Eliashberg functions, or numerical XRD spectra. For images, we have
simulated and experimental scanning transmission electron microscope (STEM) and scanning tunneling microscopy (STM) images for 2D materials. For text data, we have used the publicly available arXiv dataset.

Currently, we have models for feature-based/tabular models (such as RandomForest \cite{scikit-learn}, Gradient boosting \cite{scikit-learn}, Linear regression \cite{scikit-learn}), graph based
models (such as ALIGNN \cite{Choudhary2021_ALIGNN, gupta2024structure}, SchNet \cite{Schutt2018SchNet}, CGCNN \cite{Xie2018CGCNN}, M3GNET \cite{m3gnet}, AtomVision \cite{Choudhary2023_AtomVision}, ChemNLP \cite{ChemNLP_arxiv}) as well as transformers (such as OPT \cite{zhang2022opt}, GPT \cite{brown2020language}, and T5 \cite{raffel2020exploring}). These
models use popular AI code bases including PyTorch \cite{paszke2019pytorch}, scikit-learn \cite{scikit-learn}, TensorFlow \cite{tensorflow2015-whitepaper}, LightGBM \cite{ke2017lightgbm}, JAX \cite{jax2018github}, and HuggingFace \cite{wolf2020huggingfaces}. These models are used
for a variety of properties such as formation energies, electron bandgaps, phonon spectra, forces, text data etc. 

\subsection{Quantum computation}
Quantum chemistry is one of the most promising applications of quantum computations \cite{nielsen2001quantum}. Quantum computers with relatively few logical
qubits can potentially exceed the performance of much larger classical computers because the size of Hilbert space increases
exponentially with the number of electrons in the system. Predicting the energy levels of a Hamiltonian is a typical and
fundamentally important problem in quantum chemistry. We use Hamiltonian simulations with quantum algorithms and compare it with
classical solvers. Determination of appropriate quantum circuit for a specific QC problem is a challenging task. For example, we use the tight-binding Hamiltonians for electrons and phonons in JARVIS-DFT and evaluate the electron
bandstructures using quantum algorithms (such as variational quantum eigen solver (VQE) \cite{vqe} and variational quantum deflation
(VQD) \cite{Higgott2019variationalquantum}) and with different quantum circuits (such as PauliTwo design \cite{Qiskit} and  SU(2) \cite{Qiskit} circuits). We
primarily use the Qiskit \cite{Qiskit} software in this work through the JARVIS-Tools/AtomQC \cite{ChoudharyAtomQC} interface, but other packages such as Tequila \cite{Kottmann_2021}, Circq \cite{cirq_developers_2022_7465577},
and Pennylane \cite{bergholm2022pennylane,arrazola2023differentiable} can also be easily integrated. In addition to studying algorithm and circuit architecture dependence, the leaderboard
can be used for studying the noise-levels in quantum circuits across different quantum computers, which is a key issue hindering
quantum computer commercialization. Currently, we are only using statevector simulators for the quantum algorithms available in the Qiskit \cite{Qiskit} library.

\subsection{Experiments}

\textcolor{black}{Although experimental results for material properties and spectra are referenced in comparison to computational methods (within the JARVIS-Leaderboard and other leaderboards such as MatBench \cite{dunn2020benchmarking}), we dedicated a portion of the JARVIS-Leaderboard to experimental benchmarking. Benchmarking experiments essentially boils down to the comparison of different experiments for the same desired result/s. A systematic way to perform this benchmarking is through round-robin testing \cite{round-robin}. This is an inter-laboratory test performed independently several times, which can involve multiple scientists and a variety of methods and equipment. This approach has been applied successfully for a range of materials science applications \cite{co2-rr,4569495,rr-am,rr-nist,10.1063/1.4905250}, but many more of such experiments are still needed. Specifically in the JARVIS-Leaderboard, we include experimental round-robin results for manometric measurements of CO$_2$ adsorption \cite{co2-rr}. It is important to note that the experimental results included in the leaderboard are for well-characterized materials with well-defined properties and phenomena that can be easily reproduced (in contrast to replication attempts of variable experiments, such as the recent attempt to synthesize room temperature superconductors \cite{PhysRevB.108.235127,lee2023roomtemperature,Guo_2023,kumar2023absence}).} Some of the experiments we used for benchmarking purposes are XRD, magnetometry, vibroscopy, and scanning electron microscopy (SEM) and transition electron microscopy (TEM). We purchase the samples from industrial vendors with available identifiers such as CAS-number. We also carried out XRD for MgB$_2$ (a superconducting material) to verify its crystal structure before carrying out magnetometry measurements to determine the transition temperature. This measurement was compared with numerical XRD data. Magnetometry measurements for superconductors were also conducted to compare their superconducting transition temperatures with respect to predicted or experimentally available values \cite{doi:10.1177/0040517520918232}. Strain-stress measurements were done for Kevlar for failure analysis \cite{doi:10.1177/0040517520918232}. We have several instruments such as  Bruker D8, Titan, Quantum design PPMS and FAVIMAT in the leaderboard currently.

\subsection{Metrics used}

 We use several metrics in the leaderboard depending on the ``sub-categories'' mentioned above. We use mean absolute error (MAE),
 accuracy (acc), multi-mae \textcolor{black}{(L1 norm of multi dimensional data)},  recall-oriented understudy for gisting evaluation (ROUGE) for the singlepropertyprediction, singlepropertyclassification, spectra/eigensolver/atomic forces and textGen/textsummary
 subcategories respectively. As the user contributes their data to compare against the reference data (benchmarks), other
 complementary metrics (such as those available in the sklearn.metrics library) can be easily calculated as the raw contribution data is also made available through the website. For the sake of
 readability and ease of use, we primarily employ the metrics mentioned above. For single property prediction, there is only scalar values per column in the csv.zip file with id and prediction separate by comma., For spectra, force-prediction and other multi-value quantities (i.e.with multiple prediction values per id) we concatenate the array and separate by semicolon (to avoid comma convention in csv files). The benchmark data is also stored in a similar format. We provide tools to convert these csv.zip files into json or other file formats if needed. We also provided notebooks to visualize the data through Jupyter/Colab notebooks.  In addition, we plan to eventually add metrics for
 timing, uncertainty, development cost and other details.

\section*{Acknowledgements}

K.C., D.W., K.F.G., A.F., A.J.B., M.W., and F.T. thank the National Institute of Standards and Technology for funding, computational, and data-management
resources. This work was performed with funding from the CHIPS Metrology Program, part of CHIPS for America, National Institute of Standards and Technology, U.S. Department of Commerce. K.C. thanks the computational support from XSEDE (Extreme Science and Engineering Discovery Environment) computational resources
under allocation number TG-DMR 190095. Contributions from K.C. were supported by the financial assistance award 70NANB19H117 from the U.S.
Department of Commerce, National Institute of Standards and Technology. J.T.K., K.S., P.G. and P.R.C.K. were supported by the U.S. Department of
Energy, Office of Science, Basic Energy Sciences, Materials Sciences and Engineering Division, as part of the Computational Materials Sciences
Program and Center for Predictive Simulation of Functional Materials. A.F. and P. G. were supported by the Center for Nanophase Materials Sciences, which is a US Department of Energy, Office of Science User Facility at Oak Ridge National Laboratory.  AHR thanks the Supercomputer Center and San Diego Supercomputer Center through allocation DMR140031 from the Advanced Cyberinfrastructure Coordination Ecosystem: Services \& Support (ACCESS) program, which is supported by National Science Foundation grants \#2138259, \#2138286, \#2138307, \#2137603, and \#2138296. AHR also recognizes the support of West Virginia Research under the call Research Challenge Grand Program 2022 and NASA EPSCoR Award 80NSSC22M0173. N.M. and A.M. acknowledge support from the U.S. Department of Energy through the LANL LDRD Programs under grant no. 20210036DR and 20220814PRD4, respectively. V.G. and A.A. were supported by NIST award 70NANB19H005 and NSF award CMMI-2053929. S.H.W. especially thanks to the NSF Non-Academic Research Internships for Graduate Students (INTERN) program (CBET-1845531) for supporting part of the work in NIST under the guidance of K.C. A.M.K. acknowledges support from the School of Materials Engineering at Purdue University under startup account F.10023800.05.002. P.F. acknowledges support by the Federal Ministry of Education and Research (BMBF) under Grant No. 01DM21001B (German-Canadian Materials Acceleration Center).


Please note certain equipment, instruments, software, or materials are identified in this paper in order to specify the experimental
procedure adequately. Such identification is not intended to imply the recommendation or endorsement of any product or service by NIST,
nor is it intended to imply that the materials or equipment identified are necessarily the best available for the purpose.

This manuscript has been authored by UT-Battelle, LLC, under contract DE-AC05-00OR22725 with the US Department of Energy (DOE). The publisher acknowledges the US government license to provide public access under the DOE Public Access Plan (\url{https://www.energy.gov/doe-public-access-plan}). The Los Alamos National Laboratory is operated by the Triad National Security, LLC, for the National Nuclear Security Administration of U.S. Department of Energy (Contract No. 89233218CNA000001).

\section{Data Availability Statement}
Multiple datasets used in this work are available at the Figshare repository: \url{https://figshare.com/authors/Kamal_Choudhary/4445539}. Index and usage guidelines are provided at \url{https://pages.nist.gov/jarvis/databases/}.

 \section{Code Availability Statement}
JARVIS-Leaderboard package mentioned in the article can be found at \url{https://github.com/usnistgov/jarvis_leaderboard}.

 \section{Competing interests}
The authors declare no competing financial or non-financial interests.


 \section{Author Contributions}
K.C. conceived of the idea, developed the workflow, designed several benchmarks and oversaw the project. K.C., D.W., K.L., K.F.G. and V.G. wrote the first draft of the manuscript. K.C., D.W., K.L., K.F.G., V.G., A.H.R., J.T.K., K.S., A.F., R.W., A.M.K., K.Y., Y.L., P.R., A.M., S.H.W., E.B., A.D.R., T.D.R., A.J.B., F.T. uploaded benchmarks and contributions to the leaderboard. All authors contributed in editing and revising the manuscript. List of contributors to the GitHub repository for this work is also available at: \url{https://github.com/usnistgov/jarvis_leaderboard/graphs/contributors}.


\clearpage
\section{References}

\section{Figure legends}
1. Leaderboard snapshot with an example output for AI based formation energy per atom model on the JARVIS-DFT (dft\_3d) dataset. a) homepage sanpshot showing list of categories and number of available contributions at the time of writing, b) an example AI regression model benchmark for formation energy with several contributions. The methods are sorted based on the mean absolute error (MAE) values. Lower MAE values indicate higher accuracy, c) explicit table for the plot in panel b.Links to 
 individual csv.zip (AI-SinglePropertyPrediction-formation\_energy\_peratom-dft\_3d-test-mae.csv.zip),  json.zip (dft\_3d\_formation\_energy\_peratom.json.zip), shell script (run.sh) and detailed info (metadata.json) files are provided to help enhance reproducibility. Such results plots and tables are available for each benchmark in the leaderboard.
 
2. A tree diagram for directory and file-structure in the leaderboard. There are two main directories in the repo: (1) benchmarks (reference) and (2) leaderboard
contributions (for various leaderboard entries). In the ``benchmarks'' directory, there are folders for the AI, ES, QC, FF, and EXP categories. Within them, there are sub-folders for
specific sub-categories. In the ``contributions'' directory there is a collection of folders that consists of .csv.zip, metadata.json files, and
optionally a Dockerfile and run.sh file for available contributions from each method. The csv.zip file contains entries of identifier (id) and corresponding prediction values as
obtained by the corresponding model/method. These test identifiers (such as JVASP-1408)  must match the test set ids in the
json.zip file in the benchmarks folder for the metric measurements to work.

3. A tree diagram for directory and file-structure in the leaderboard. There are two main directories in the repo: (1) benchmarks (reference) and (2) leaderboard
contributions (for various leaderboard entries). In the ``benchmarks'' directory, there are folders for the AI, ES, QC, FF, and EXP categories. Within them, there are sub-folders for
specific sub-categories. In the ``contributions'' directory there is a collection of folders that consists of .csv.zip, metadata.json files, and
optionally a Dockerfile and run.sh file for available contributions from each method. The csv.zip file contains entries of identifier (id) and corresponding prediction values as
obtained by the corresponding model/method. These test identifiers (such as JVASP-1408)  must match the test set ids in the
json.zip file in the benchmarks folder for the metric measurements to work.

4. Distribution of data in each dataset. (a) all entries in leaderboard, (b) entries with unique identifiers. Note that one identifier (such as JVASP-1002 for silicon) can have multiple properties (such as bandgap, bulk modulus etc.). A script to generate this figure is also provided on the leaderboard website as the leaderboard is continuously evolving.

5. Periodic table element distribution for entries in all the datasets. This is calculated by taking into account all the element specific entries normalized by total entries i.e. these are percentage probabilities.

6. Example mean absolute errors for benchmarks including (a) artificial intelligence (AI) formation
energy for test set with 5572 materials in JARVIS-DFT 3D dataset, (b) electronic structure (ES) Si (JARVIS-DFT ID: JVASP-1002) bandgap, (c) classical force-field (FF) based Voigt bulk modulus of Si and (d) machine learning force-field (MLFF) based forces for Si. We provide Jupyter/Google colab notebooks to easily plot such comparisons for all available benchmarks. Also, similar analysis figures for all the available benchmarks are available in the supplementary information (Supplementary Figures 1-298). As a note, these plots are a current snapshot of the leaderboard, and it is possible that new and more accurate models will be developed and added here in the future.

7. Relative performance computed as the ratio of the MAE difference between the best descriptor-based model and the best neural networks to the MAE of the best neural networks in the AI regression benchmarks. The benchmark name and the corresponding best performing neural network are indicated in the left and right y axis, respectively. For all the considered AI benchmarks, the best descriptor-based model is the tree-based model using Magpie \cite{magpie} and Voronoi-tessellation \cite{ward2017including} features. As a disclaimer, these plots are a current snapshot of the leaderboard, and it is possible that new and more accurate models will be developed in the future.

8. Mean absolute deviation (MAD) to mean absolute error (MAE) ratio for (a) AI and (b) electronic structure methods. MAD:MAE serves as uniform criteria for comparing performances of models.

9. Example results for AI, ES, QC and EXP results. (a) formation-energy-per atom model using AI for JARVIS-DFT 3D dataset with 5572 materials in the test set, (b) bulk modulus predictions using ES methods for 21 materials, (c) electronic bandstructure of Aluminum using QC methods with different quantum circuits on a coarse k-point mesh, (d) CO$_2$ capture for zeolite (ZSM-5) at several labs in inter-laboratory/round-robin fashion.

\section{Table legends}

1. Comparison of benchmark infrastructure available for materials design methods for several categories.

2. Summary of current benchmark categories and methods available in the JARVIS-Leaderboard at the time of writing. More details can be found in the individual metadata.json file. Note that the number of methods is continuously growing.



\end{document}